\documentclass[10pt, a4paper]{article}
\pdfoutput=1
\usepackage{jcappub}
\usepackage{amsmath,amssymb,graphicx,xspace,subfigure,tikz}
\usepackage{xfrac}	    
\usepackage{bm}
\usepackage{mathrsfs}
\usepackage{ulem}
\usepackage{hyperref} 
\usepackage{enumitem}
\hypersetup{
    colorlinks=true,       
    linkcolor=red,          
    citecolor=blue,        
    filecolor=magenta,      
    urlcolor=blue           
}
\usepackage[all]{hypcap}

\newcommand{\dd}{\mathrm{d}}
\newcommand{\ee}{\mathrm{e}}

\newcommand{\xvec}{\vec{x}}
\newcommand{\pvec}{\vec{p}}
\newcommand{\BBN}[1]{\text{\sc bbn}}
\newcommand{\PBH}[1]{\text{\sc pbh}}
\newcommand{\SM}[1]{\text{\sc sm}}
\newcommand{\CGPP}[1]{\text{$\mathrm{CGPP}$}}
\newcommand{\eref}[1]{eq.~(\ref{#1})}

\newcommand{\erefs}[2]{eqs.~(\ref{#1})~and~(\ref{#2})}

\newcommand{\MPl}{M_\text{Pl}}
\newcommand{\ssfrac}[2]{\sfrac{#1\mkern-2.2mu}{#2}}

\newcommand{\bes}[1]{\begin{equation}\begin{split} #1 \end{split}\end{equation}}

\newcommand{\ie}[1]{\text{i.e.}}
\newcommand{\eg}[1]{\text{e.g.}}

\newcommand{\fref}[1]{fig.~\ref{#1}}
\newcommand{\Fref}[1]{Fig.~\ref{#1}}

\newcommand{\sref}[1]{sec.~\ref{#1}}

\renewcommand{\eref}[1]{eq.~(\ref{#1})}

\newcommand{\rref}[1]{ref.~\cite{#1}}

\newcommand{\undert}[2]{{\color{blue}{\underbrace{\color{black}{#1}}_{#2}}}}

\title{\begin{center}Setting up stasis \\ with gravitational interactions\end{center}}

\author[a]{Andrew J. Long,}
\emailAdd{andrewjlong@rice.edu}

\author[b]{\ Barmak Shams Es Haghi,}
\emailAdd{shams@austin.utexas.edu}

\author[a]{ and Moira Venegas}
\emailAdd{mv58@rice.edu}

\affiliation[a]{Department of Physics and Astronomy, Rice University, Houston, Texas 77005, U.S.A.}
\affiliation[b]{Texas Center for Cosmology and Astroparticle Physics, Weinberg Institute for Theoretical Physics, Department of Physics, The University of Texas at Austin, Austin, Texas 78712, U.S.A.}

\abstract{
An epoch known as cosmological stasis may have taken place in the early Universe.  
During matter-radiation stasis, a population of non-relativistic particles with different masses gradually decay into relativistic particles, and the effective equation of state $w$ remains approximately constant at a value between that of matter ($w=0$) and that of radiation ($w=\ssfrac{1}{3}$).  
In this work, we investigate how to set up the appropriate initial conditions for stasis using gravitational interactions.  
We consider two scenarios: that the tower of non-relativistic particles is populated by the evaporation of primordial black holes (PBHs) and that the tower is populated by cosmological gravitational particle production (CGPP) during inflation.  
We calculate the abundance of particles on different levels of the tower to assess whether stasis is viable. 
We find that both scenarios can provide the needed initial conditions for stasis, and that they predict distinctive scaling exponents $\Omega_l \propto m_l^\alpha$ with mass $m_l$. 
}

\begin{document}

\maketitle

\section{Introduction}
\label{sec:intro}
Cosmology textbooks describe an epoch of inflaton domination, which gives way to an epoch of radiation domination, followed by an epoch of dark matter domination, and finally an epoch of dark energy domination today~\cite{Baumann:2022mni}.  
This is a natural progression.  
As a consequence of the cosmological expansion, the energy density carried by each component decreases with time, going as $\rho(t) \propto a(t)^{-3(1+w)}$ where $a(t)$ is the Friedmann-Robertson-Walker (FRW) scale factor at time $t$ and $w$ is the equation of state~\cite{Kolb:1981hk}.  
For two components with different equations of state, such as radiation ($w=1/3$) and matter ($w=0$), their energy densities dilute at different rates.  
As a result, an initial epoch of radiation domination naturally gives way to an epoch of matter domination.  
Between these epochs, the `transfer of power' occurs quickly in just a Hubble time as $a(t)$ grows by an order-one factor. 
In other words, there is no sustained period of time where different components share fixed fractions of the energy budget.  

This standard cosmological timeline can arise when there is no significant interaction among the various components.  
However, if there is a channel for energy transfer from one component to another, then they can evolve together with a common equation of state, even for many Hubble times.\footnote{This behavior was observed by the authors of \rref{Long:2015cza} who studied plasma oscillations with magnetic monopoles.  The universe remains matter dominated, but the matter's kinetic energy is in stasis with the magnetic field's energy.  }    
This idea was proposed by the authors of \rref{Dienes:2021woi}, who gave it the name ``cosmological stasis,'' and in the last few years, stasis has been explored in various contexts~\cite{Dienes:2022zgd,Dienes:2023ziv,Dienes:2024wnu,Barber:2024vui,Barber:2024izt,Halverson:2024oir,Dienes:2025tox,Huang:2025odd}.  
From the experimental perspective, it is important to assess the possible signatures of stasis in current and future cosmological observables~\cite{Batell:2024dsi}.  
From the theoretical perspective, it is useful to ensure that the assumed ingredients that lead to stasis are well-motivated and grounded.  
To that end, in this work we investigate whether minimal gravitational interactions are sufficient to set up the necessary initial conditions for stasis.  

We briefly review the necessary ingredients for matter-radiation stasis~\cite{Dienes:2021woi}.  
Consider a tower of $N$ particle species indexed by a level number $l \in \{ 1, 2, \cdots, N \}$.  
The particles on level $l$ are assumed to have mass $m_l$ given by the power-law relation: 
\begin{equation}\label{eq:m_l}
    m_l = m_1 + (l-1)^\delta \, \Delta m
    \qquad \text{for $l = 1, 2, \cdots, N$} 
    \;.
\end{equation}
We take $m_1 \geq 0$, $\Delta m > 0$, and the index $\delta > 0$ as free parameters. 
It follows level $l=1$ is the lightest, and level $l=N$ is the heaviest, \ie{} $m_1 < m_N$.  
For our numerical work we focus on models with $\delta = 1$ and $m_1 = \Delta m \gg 100 \, \mathrm{GeV}$, such that the whole tower is heavier than the Standard Model.  

All of the $N$ particle species are assumed to be unstable and decay outside the tower to an additional (radiation) sector consisting of light particles ($\ll m_1$).  
The particles on level $l$ are assumed to have decay rate $\Gamma_l$ given by the power-law relation: 
\begin{equation}\label{eq:Gamma_l}
    \Gamma_l = \left(\frac{m_l}{m_1}\right)^{\!\!\gamma} \Gamma_1 
    \;.
\end{equation}
We take $\Gamma_1 > 0$ and the index $\gamma$ as free parameters.  
For our numerical work we focus on models with $\gamma = 5$ and $\Gamma_1 = 10^{-40} \MPl$.  
This hierarchy of decay rates is essential for achieving successful stasis.  
The heaviest particles decay first, thereby effectively transforming from matter into radiation.  
Afterward, each lighter species `takes a turn' at being the dominant matter component before passing its energy into radiation.  
The stasis phase lasts from $t_S = \Gamma_N^{-1}$ when the heaviest particle decays until $t_R = \Gamma_1^{-1}$ when the lightest particle decays and the system returns to radiation domination~\cite{Dienes:2021woi}. 

The final key ingredient for successful stasis is a correct population of particles in the tower.  
The work on cosmological stasis assumes that the particles in the tower are non-relativistic and that they have initial abundances following a power-law relation: 
\begin{equation}\label{eq:Omegal_stasis}
    \Omega_l = \left(\frac{m_l}{m_1}\right)^{\!\!\alpha} \Omega_1 
    \qquad \text{(initially)} 
    \;.
\end{equation}
Here $\Omega_l = \rho_l / \rho_\mathrm{tower}$ is the cosmological energy fraction carried by particles of species $l$ with energy density $\rho_l \approx m_l n_l$, and where $\rho_\mathrm{tower} = \sum_l \rho_l$ is the energy density carried by the entire tower.  

Stasis is achieved when the inevitable dominance of tower of particles (matter) over radiation in an expanding Universe is counterbalanced by the gradual decay of the tower into radiation. 
This balance leads to an extended period during which the abundances of both matter and radiation, $\Omega_M$ and $\Omega_R$, remain constant. 
Stasis does not necessarily correspond to a matter- or radiation-dominated regime; instead, the ratio of abundances, $\Omega_M / \Omega_R$, approximately remains constant, which results in an effective equation of state $w_S = w_R \Omega_R + w_M \Omega_M \approx \Omega_R / 3$ with $0 \lesssim  w_S \lesssim 1/3$.
Not every value of $\alpha$ allows for an extended phase of stasis.  
In order for stasis to occur, the three exponents $\delta$, $\gamma$, and $\alpha$ must satisfy the inequalities~\cite{Dienes:2021woi}:
\begin{equation}\label{eq:constraint}
   - \frac{1}{\delta} <\alpha\leq \frac{\gamma}{2}-\frac{1}{\delta}
    \;.
\end{equation}
It has been shown that for a tower of particles parameterized by exponents $\delta$, $\gamma$, and $\alpha$, the abundances during stasis are given by~\cite{Dienes:2021woi}:
\begin{equation}
    \Omega_M = \frac{2\gamma\delta-4(1+\alpha\delta)}{2\gamma\delta-(1+\alpha\delta)}
    \quad \text{and} \quad 
    \Omega_R = \frac{3(1+\alpha\delta)}{2\gamma\delta-(1+\alpha\delta)}
    \;.
\end{equation}
In the next paragraph we motivate the choices $\delta = 1$ and $\gamma = 5$, which imply $-1 < \alpha \leq 3/2$.  

Certain choices for these power-law relations arise naturally in theoretically-compelling theories of new physics. 
For example, the Kaluza-Klein excitations~\cite{Bailin:1987jd} of a five-dimensional scalar field $\phi(\xvec,\theta,t)$ that is compactified on a circle of radius $R$ have a mass spectrum given by $m_l = \sqrt{m^2 + (2 \pi l/R)^2}$, where $m$ denotes the four-dimensional scalar mass.  
If $m R \ll 1$ then $m_l \approx m$, and if $m R \gg 1$ then $m_l \approx (2 \pi/R) l$.  
Collectively the two limiting regimes are captured by $m_l \approx m + (2\pi/R) l$, which takes the same form as \eref{eq:m_l} with $m_1 = m$, $\Delta m = 2\pi/R$, and $\delta = 1$.  
This observation motivates our choice of $\delta = 1$ and $m_1 = \Delta m \gg 100 \, \mathrm{GeV}$, which we adopt for numerical examples.
While theories with compactified dimensions predict a tower of states with power-law relations, we use these as motivation but focus our analysis on gravitational particle production in the absence of such extra dimensions. 
As for the decay rates of particles in the tower, if the scalar field $\phi_l$ decays through a dimension-$d$ operator into radiation, then $\gamma = 2d-7$.  
For example, $\gamma=5$ can be realized from a $d=6$ operator that couples the scalars to the Weinberg operator, \ie{}, $\mathscr{L} = c_l \phi_l (LH) (LH) / \Lambda^2$ where $L$ is the lepton doublet, $H$ is the Higgs doublet, and $\Lambda$ is the ultraviolet cutoff.  
The decay rate $\Gamma_l \sim m_l^5 / \Lambda^4$ follows from dimensional analysis, and it corresponds to $\gamma = 5$ in \eref{eq:Gamma_l}.  
For example, if $m_1 \sim 10^{-7} \MPl$ and $\Lambda \sim \MPl$ then $\Gamma_1 \sim m_1^5 / \Lambda^4 \sim 10^{-35} \MPl$.  
We use similar parameter values for our numerical examples. 
The motivated choices of $\delta$ and $\gamma$, together with the predicted values of $\alpha$ from our study based on populating the tower via gravitational interactions, are found to be compatible with the stasis constraint. 

The central question that we seek to address in this work is whether or not the initial condition for stasis results from only the gravitational interactions of the particles in the tower.  
Namely, does the relic abundance have a power-law relationship with the mass $\Omega_l \propto m_l^\alpha$ as in \eref{eq:Omegal_stasis}, and if so, does the exponent $\alpha$ obey \eref{eq:constraint}? 
There are two notable mechanisms of particle production through gravitational interactions.  
First, the evaporation of black holes via Hawking radiation leads to particle creation, and in this work we investigate whether a population of primordial black holes (PBHs) can give rise to \eref{eq:Omegal_stasis}.\footnote{PBHs have previously been studied in the context of stasis~\cite{Dienes:2022zgd}, where they have an extended mass function, act as non-relativistic matter that dominates the energy density of the Universe and evaporates into radiation. In this study, PBHs emit heavy, unstable particles that eventually decay and drive the Universe toward stasis. As we show, in our set-up, PBHs neither need to dominate the Universe nor have an extended mass function in order to lead to stasis. 
} 
Second, cosmological gravitational particle production (CGPP) during inflation and at the end of inflation leads to particle creation, and we briefly discuss its implications for stasis. 

As we show in this paper, when PBHs populate the tower, the relic abundances of the emitted particles follow a power-law relationship, the exponent of which differs depending on whether the particle mass is lighter or heavier than the initial temperature of the PBHs. Particles lighter than the temperature of the PBHs are ultra-relativistic upon emission and require cosmological redshifting to become non-relativistic, while heavier ones are semi-relativistic and become non-relativistic quickly. In general, depending on the states in the tower and the mass of the PBHs, one would expect a combination of these two possibilities. For CGPP  during inflation, we also find power-law relationships for abundances of particles with exponents that depend on the particle spin. Based on particle masses, the inflationary scale, and the reheating temperature, CGPP can populate a tower of states with non-relativistic particles that may come to dominate the Universe. It is worth mentioning that the exponents associated with PBHs evaporation are distinct from those arising from CGPP.

The outline of this paper is as follows. 
In \sref{sec:PBH_evap}, after reviewing the Hawking evaporation of PBHs, we analytically calculate the abundances of emitted particles and validate our results numerically by using a Boltzmann equation formalism. 
We examine how the spectra of emitted particles evolve over the lifetime of the black hole and highlight distinctive features in the final spectrum. 
In the final part of this section, by allowing the particles emitted by PBHs to decay, we discuss the possibility that these decays could set the Universe on a trajectory toward stasis. 
In \sref{sec:gpp}, we consider CGPP during inflation as an alternative way of preparing the necessary initial conditions for stasis, relying on gravitational interactions. 
Finally, we discuss our findings and conclude in \sref{sec:conclusion}.

\section{Primordial black hole evaporation}
\label{sec:PBH_evap}

In this section we explore the idea that an evaporating population of PBHs creates the tower of particles and thereby sets the initial condition needed for stasis.  
There have been numerous studies investigating the production of cosmological relics through PBH evaporation~\cite{Fujita:2014hha,Morrison:2018xla,Lennon:2017tqq,Allahverdi:2017sks,Hooper:2019gtx,Masina:2020xhk,Baldes:2020nuv,Gondolo:2020uqv,Auffinger:2020afu,Bernal:2020bjf,Arbey:2021ysg,Sandick:2021gew,Cheek:2021odj,Gehrman:2022imk,Gehrman:2023esa,Gehrman:2023qjn,Shallue:2024hqe}, and we draw upon this body of literature.

\subsection*{Black hole evaporation}
\label{sub:PBH_evap}

A black hole is expected to lose mass by emitting particles via Hawking radiation~\cite{Hawking:1974rv,Hawking:1975vcx}, and when all of its mass is exhausted, the black hole is said to have evaporated completely.  
We consider a hypothetical population of PBHs that may have existed in the early universe~\cite{Hawking:1971ei,Carr:1974nx,Carr:1975qj}, and we are interested in the products of their evaporation.  

To study PBH evaporation, we make a few simplifying assumptions.  
First we suppose that all the PBHs have the same mass at the time of formation.  
It is straightforward to generalize our analysis to account for a broader mass distribution.  
Second we consider black holes that have negligible angular momentum and electromagnetic charge.  
It is straightforward to extend our analysis to account for radiation from spinning or charged black holes~\cite{Page:1976df,Page:1976ki,Page:1977um}. 
Third we use geometric optics to approximate the greybody factors~\cite{Page:1976df}, which modulate the spectrum so as to account for the probability that a radiated particle escapes the black hole's gravitational well and reaches spatial infinity.
It is straightforward to go beyond geometrical optics and include these spin-dependent factors that suppress emission for low-momentum particles, but they would not significantly impact the total number of emitted particles, which is related to our observable of interest. 

Under these assumptions, we approximate Hawking radiation as black body spectrum where 
\bes{\label{eq:TPBH_from_MPBH}
    T_\PBH{}(t) = \frac{\MPl^2}{8\pi M_\PBH{}(t)}
    \;,
}
is the instantaneous horizon temperature of the black hole at a time $t$ when the black hole mass is $M_\PBH{}(t)$.  
We denote the Planck mass by $\MPl = 1 / \sqrt{G} \approx 1.22 \times 10^{19} \, \mathrm{GeV}$, and we set $\hbar = c = 1$.  

The energy carried away by Hawking radiation is lost from the mass of the black hole.  
Consequently, the black hole mass $M_\PBH{}(t)$ decreases according to 
\bes{\label{eq:dMPBHdt}
    \frac{\dd}{\dd t} M_\PBH{} = - \sum_\text{species} P(t) = -\frac{9}{40960 \pi} \, g_\ast(t) \frac{\MPl^4}{M_\PBH{}(t)^2}
    \;.
}
The sum of the rate of energy emission $P(t)$ runs over all particle species in the theory, and the factor $g_\ast(t)$ counts the effective number of light degrees of freedom emitted by the black hole at time $t$ (see below).  
If a black hole has initial mass $M_0$ when it was formed at time $t = t_0$, then its mass $M_\PBH{}(t)$ at later time $t > t_0$ is calculated by solving \eref{eq:dMPBHdt}.  
If $g_\ast(t)$ changes slowly, then the solution is approximately 
\bes{\label{eq:MPBH}
    M_\PBH{}(t) 
    = M_0 \biggl( 1 - \frac{t - t_0}{\tau_\PBH{}(t)} \biggr)^{\!\!1/3}
    \qquad \text{where} \qquad 
    \tau_\PBH{}(t) = \frac{40960\pi}{27} \frac{1}{g_\ast(t)} \frac{M_0^3}{\MPl^4} 
    \;.
}
The black hole reaches zero mass after a time $\tau_\PBH{}$ has elapsed, and therefore $\tau_\PBH{}$ is interpreted as the black hole lifetime. 
A black hole with a larger initial mass $M_0$ takes longer to fully evaporate since $\tau_\PBH{} \propto M_0^3$.  
The black hole's Schwartzchild radius $R_S(t)$ at time $t$ is given by 
\bes{\label{eq:RS}
    R_S(t) 
    = \frac{2 M_\PBH{}(t)}{\MPl^2} 
    = R_0 \biggl( 1 - \frac{t - t_0}{\tau_\PBH{}(t)} \biggr)^{\!\!1/3}
    \qquad \text{where} \qquad 
    R_0 = \frac{2 M_0}{\MPl^2} 
    \;,
}
and the black hole shrinks to zero size as it evaporates. 
Combining \erefs{eq:TPBH_from_MPBH}{eq:MPBH} gives 
\bes{\label{eq:TPBH}
    T_\PBH{}(t) = T_0 \biggl( 1 - \frac{t - t_0}{\tau_\PBH{}(t)} \biggr)^{\!\!-1/3}
    \qquad \text{where} \qquad 
    T_0 = \frac{\MPl^2}{8\pi M_0}\;,
}
is the initial black hole temperature.  
As the black hole evolves toward smaller mass, its temperature grows without bound and formally diverges at time $t = t_0 + \tau_\PBH{}$. 

Of course the calculations presented here break down when $M_\PBH{}(t)$ reaches the scale of quantum gravity, where the effective field theory description is no longer applicable.  
Imposing $M_\PBH{} > \MPl$ implies $T_\PBH{} < \MPl / 8\pi \approx 4.85 \times 10^{17} \, \mathrm{GeV}$.  
We identify the black hole evaporation time $t_\mathrm{ev}$ to be the time such that $M_\PBH{}(t_\mathrm{ev}) = \MPl$, and this condition resolves to 
\bes{\label{eq:t_ev}
    t_\mathrm{ev} = t_0 + \biggl( 1 - \frac{\MPl^3}{M_0^3} \biggr) \, \tau_\PBH{}(t_\mathrm{ev}) 
    \;,
}
which gives $t_\mathrm{ev}$ to be very slightly smaller than $t_0 + \tau_\PBH{}$. 
The distinction between $t_\mathrm{ev}$ and $\tau_\PBH{}$ will only play a role in our discussion of the very high-momentum tail of the spectrum of emitted particles. 

\subsection*{Hawking radiation spectrum}
\label{sub:radiation}

Consider particles of species $l$, which have mass $m_l$, internal degrees of freedom $g_l$ (\eg{}, color, spin), and quantum statistics indicator $s_l = -1$ for bosons or $s_l = +1$ for fermions.  
A black hole with temperature $T_\PBH{}(t)$ at time $t$ emits particles of species $l$ provided that $m_l$ is not much larger than $T_\PBH{}(t)$.  
The average number of particles of species $l$ that are emitted with magnitude of momentum between $p$ and $p + \dd p$, between the time $t$ and $t + \dd t$ is given by 
\bes{\label{eq:dNl}
    \dd N_l = \frac{g_l}{8 \pi^2} \, \frac{p^3}{E_l(p)} \, \frac{1}{\ee^{E_l(p) / T_\PBH{}} + s_l} \, \bigl( 4\pi b_S^2 \bigr) \, \dd p \, \dd t 
    \;,
}
where $E_l(p) = (p^2 + m_l^2)^{1/2}$ is the energy of the emitted particle far from the black hole.  
Since the emission is homogeneous across the black hole's surface, integrating over the whole horizon yields the factor of $4 \pi b_S^2 = \tfrac{27}{4} 4 \pi R_S^2$, which is the geometrical optics approximation to the greybody factor. 
The effective radius $b_S = 3\sqrt{3} R_S/2$ is slightly larger than the Schwartzchild radius $R_S$, since it corresponds to the impact parameter at which a massless particle will precisely fall into the circular orbit around the black hole. 
The rate of particle emission $\dot{N}_l(t)$ at time $t$ is calculated by integrating over the momentum, which gives 
\bes{\label{eq:dotNl}
    \dot{N}_l(t) 
    = \int_0^\infty \! \dd p \, \frac{\dd N_l}{\dd t} 
    & = \frac{27 \zeta(3)}{512 \pi^4} \, g_l f_N(m_l / T_\PBH{}) \, \frac{\MPl^2}{M_\PBH{}}\;, 
}
for a bosonic species ($s_l = -1$), and the result is smaller by a factor of approximately $3/4$ for a fermionic species ($s_l = 1$).  
Similarly the rate of energy emission $P_l(t)$ at time $t$ is calculated as 
\bes{\label{eq:Pl}
    P_l(t) 
    = \int_0^\infty \! \dd p \, E_l(p) \, \frac{\dd N_l}{\dd t} 
    & = \frac{9}{40960 \pi} \, g_l f_E(m_l / T_\PBH{}) \, \frac{\MPl^4}{M_\PBH{}^2}\;, 
}
for a bosonic species ($s_l = -1$), and the result is smaller by a factor of approximately $7/8$ for a fermionic species ($s_l = 1$).  
The dimensionless functions $f_N(x)$ and $f_E(x)$ approach $1$ as $x \to 0$ and they approach $0$ exponentially quickly $\propto \ee^{-x}$ for $x \gtrsim 1$.  
The power $P_l(t)$ appears in \eref{eq:dMPBHdt}, which governs the evolution of $M_\PBH{}(t)$, and it gives rise to the factor $g_\ast(t)$.  
Assuming that the theory consists of the Standard Model particles plus a tower of $N$ additional bosonic particle species, then we can write 
\bes{\label{eq:gast}
    g_\ast(t) = g_\SM{} + \sum_{l=1}^N g_l f_E(m_l/T_\PBH{}(t))
    \;,
}
where $g_\SM{} = 106.75$ counts the Standard Model particles, which are assumed to be relativistic, \ie{} $T_\PBH{}(t) \geq T_0 > 100 \, \mathrm{GeV}$ at all times.

\subsection*{Cosmological PBH population}
\label{sub:cosmo_PBH_pop}

We assume that PBH formation, radiation, and evaporation all occur in the very early universe.  
We assume that the cosmological energy budget is initially dominated by a plasma of relativistic particles (\ie{}, radiation-dominated universe) with temperature $T(t)$ at time $t$.  
The energy density of this plasma and the Hubble parameter at time $t$ are given by~\cite{Kolb:1981hk} 
\begin{equation}\label{eq:rho_rad}
    \rho_\mathrm{rad}(t) = \frac{\pi^2}{30} \, g_E(t) \, T(t)^4 
    \qquad \text{and} \qquad 
    H(t) = \sqrt{\frac{8 \pi \rho_\mathrm{rad}(t)}{3 \MPl^2}} = \frac{1}{2t}
    \;,
\end{equation}
where $g_E(t)$ is the effective number of relativistic degrees of freedom.  
As the universe expands the cosmological scale factor $a(t)$ grows larger, and the plasma temperature $T(t)$ decreases.  
If the expansion is adiabatic, then the plasma's entropy density $s_\mathrm{rad}(t)$ decreases where~\cite{Kolb:1981hk} 
\begin{equation}
    s_\mathrm{rad}(t) = \frac{2\pi^2}{45} \, g_S(t) \, T(t)^3 
    \qquad \text{such that} \qquad 
    a(t)^3 s_\mathrm{rad}(t) = \mathrm{constant} 
    \;,
\end{equation}
and $g_S(t)$ is the effective number of relativistic degrees of freedom at time $t$.  
Since black hole evaporation produces additional radiation, it disrupts the scaling $s_\mathrm{rad} \propto a^{-3}$, but the effect is negligible provided that the universe is radiation-dominated at the time of black hole evaporation; then the entropy injection is tiny compared with the entropy of the plasma. 

If PBH formation takes place (at time $t_0$) in a radiation-dominated universe due to large overdensities entering the cosmological horizon, then the PBH mass is typically on the order of the total mass contained within a Hubble volume (\ie{}, the Hubble mass).  
We assume that all PBHs have the same mass, which we take to be \cite{Carr:1974nx,Carr:1975qj} 
\begin{equation}\label{eq:initialmass}
    M_0 = \frac{4\pi}{3} w^{3/2} \, \rho_\mathrm{rad}(t_0) \, H(t_0)^{-3}
    \;,
\end{equation}
where the numerical factor $w^{3/2} \approx 0.192$ is related to the equation of state of the cosmological medium ($w = 1/3$ for radiation-dominated universe).  
At the time of formation we denote the initial number density of PBHs as $n_{\PBH{},0}$ and the initial energy density carried by PBHs as $\rho_{\PBH{},0}$.  
It is customary to write 
\begin{equation}\label{eq:beta}
    \rho_{\PBH{},0} = \beta \, \rho_\text{rad}(t_0)
    \;,
\end{equation}
which defines the dimensionless parameter $\beta$ as the fraction of the cosmological energy budget carried by PBHs at their time of formation.   
We remain agnostic to the mechanism of PBH formation, and we take $\beta \in [0,1)$ as a free parameter.  

After PBH formation has finished, we denote the number density of PBHs as $n_\PBH{}(t)$ and the energy density carried by PBHs as $\rho_\PBH{}(t)$ at time $t$.  
Assuming that the PBHs are non-relativistic (in the cosmological rest frame), then the energy that they carry is just their rest energy.  
We further assume that the PBHs do not merge with one another, nor do they accrete an appreciable fraction of their total mass.  
Under these assumptions, the energy density and number density are related by 
\bes{\label{eq:rhoPBH}
    \rho_\PBH{}(t) = M_\PBH{}(t) \, n_\PBH{}(t) 
    \;,
}
where $M_\PBH{}(t)$ is mass of a freely-decaying PBH from \eref{eq:MPBH}. 
Due to the cosmological expansion, the scale factor $a(t)$ grows, and the densities decrease.  
The cosmological number density decreases as 
\bes{\label{eq:nPBH}
    n_\PBH{}(t) = n_{\PBH{},0} \, \biggl( \frac{a(t)}{a_0} \biggr)^{\!\!-3} 
    \;,
}
where $a_0 = a(t_0)$ is the scale factor at the time of PBH formation. 

If the initial PBH abundance is not too small, then the PBHs will come to dominate the cosmological energy budget before they decay. 
Since $n_\PBH{}(t) \propto a(t)^{-3}$ the PBH energy density scales as $\rho_\PBH{}(t) \propto a(t)^{-3}$ on time scales short compared to the PBH lifetime ($t - t_0 \ll \tau_\PBH{}$).  
Meanwhile, the energy density carried by the plasma of relativistic particles scales as $\rho_\mathrm{rad}(t) \propto T(t)^4 \propto a(t)^{-4}$.  
Even if the PBHs are a subdominant component of the energy budget at the time of their formation ($\beta \ll 1$), they may eventually come to dominate after sufficient cosmological redshifting has occurred.  
Let $t_\PBH{}$ denote the time at which the PBH population first dominates the cosmological energy budget (\ie{}, PBH domination).  
It is given by 
\bes{\label{eq:tPBH}
    & t_\PBH{} = t_0 / \beta^2 
    \qquad \text{when} \qquad 
    \rho_\PBH{}(t_\PBH{}) = \rho_\mathrm{rad}(t_\PBH{}) 
    \qquad \text{and} \qquad 
    a(t_\PBH{}) = a(t_0) / \beta 
    \\ & \qquad \qquad 
    \qquad \text{and} \qquad 
    T(t_\PBH{}) = \biggl( \frac{g_S(t_0)}{g_S(t_\PBH{})} \biggr)^{\!\!1/3} \, \beta \, T(t_0) 
    \;.
}
Note that $T(t_0) \neq T_0$ since $T_0 \equiv T_\PBH{}(t_0)$ denotes the initial PBH temperature, whereas $T(t_0)$ denotes the temperature of the ambient radiation. 
A smaller abundance parameter $\beta$ delays the onset of PBH domination.  
We focus on the regime (note: $1 \, \mathrm{gram} \approx 4.5947 \times 10^4 \MPl$) 
\begin{equation}\label{eq:beta_vs_betac}
    \beta < \beta_c \equiv 
    \biggl( \frac{27 g_\ast}{40960 \pi w^{3/2}} \biggr)^{\!\!1/2} 
    \frac{\MPl}{M_0} 
    \approx 
    \bigl( 7.4 \times 10^{-6} \bigr) 
    \biggl( \frac{g_\ast}{106.75} \biggr)^{\!\!1/2} 
    \biggl( \frac{M_0}{1\;\mathrm{gram}} \biggr)^{\!\!-1} 
    \;,
\end{equation}
such that $t_\mathrm{ev} < t_\PBH{}$ and the black holes evaporate in the radiation era, before they come to dominate.

\subsection*{Relic abundance estimates}
\label{sub:estimates}

In this section we make several simplifying assumptions in order to derive approximate expressions for the initial relic abundance of particles produced from PBH evaporation. 
In the following sections we validate these approximations by a direct numerical integration of the Boltzmann equation.  

At this point it is useful to introduce the distinction between light and heavy particles.  
We say that a particle with mass $m_l$ is light at time $t$ if $m_l < T_\PBH{}(t)$, and we say that a particle is heavy at time $t$ if $m_l > T_\PBH{}(t)$.  
As a black hole evaporates, its mass decreases according to \eref{eq:MPBH} and its temperature increases according to \eref{eq:TPBH}.  
So particles with $m_l < T_0$ are light when PBH begins to evaporate, and they always remain light.  
On the other hand, particles with $m_l > T_0$ are heavy when PBH starts evaporation, but they eventually become light at the time 
\begin{equation}\label{eq:tm}
    t_m = t_0 + \biggl( 1 - \frac{T_0^3}{m_l^3} \biggr) \tau_\PBH{}(t_m) 
    \qquad \text{when} \qquad 
    m_l = T_\PBH{}(t) 
    \;.
\end{equation}
Even particles that are initially very heavy, $m_l \gg T_0$, will eventually become light, albeit possibly only briefly during the final stage of black hole evaporation. 
Particles that are heavy at time $t$ can not be emitted efficiently from a black hole, because of the kinematic mismatch, and this behavior is captured by the Boltzmann suppression factor $\propto \ee^{-m_l / T_\PBH{}}$ arising from \eref{eq:dNl}.  
This observation motivates a simplifying approximation, 
\bes{
    f_N(m_l/T_\PBH{}) \approx f_E(m_l/T_\PBH{}) \approx \begin{cases} 
    1 & , \quad m_l < T_\PBH{}(t) \quad \text{(light at time $t$)} \\ 
    0 & , \quad m_l > T_\PBH{}(t) \quad \text{(heavy at time $t$)} 
    \end{cases}
    \;,
}
which we employ in order to derive analytical results.    
Particles that are heavy at time $t$ are approximated to have zero emission.  
Note that this approximation allows \eref{eq:gast} to be written as 
\bes{\label{eq:gast_approx}
    g_\ast = 
    \begin{cases} 
    g_\SM{} + N & , \quad m_1 < m_N < T_0 \quad \text{(whole tower always light)} \\  
    g_\SM{} & , \quad T_0 < m_1 < m_N \quad \text{(whole tower initially heavy)}
    \end{cases}
    \;,
}
To illustrate the utility of this approximation, we use it to calculate the total number of particles emitted over the lifetime of the black hole.  
This quantity is obtained by integrating $\dot{N}_l$ from $t = t_0$ to $t_\mathrm{ev} \approx t_0 + \tau_\PBH{}$.  
For a bosonic species ($s_l = -1$), the result is 
\begin{equation}\label{eq:Nl_cases}
    N_l = \begin{cases}
    \frac{15 \zeta(3)}{\pi^4} \frac{g_l}{g_\ast} \frac{M_0}{T_0}
    & , \quad m_l < T_0 \quad \text{(always light)} \\ 
    \frac{15 \zeta(3)}{\pi^4} \frac{g_l}{g_\ast} \frac{M_0 T_0}{m_l^2}
    & , \quad m_l > T_0 \quad \text{(initially heavy)}
    \end{cases}
    \;,
\end{equation}
and for a fermionic species ($s_l = 1$) the result is smaller by a factor of $3/4$.  
Recall that \eref{eq:TPBH} relates $M_0$ and $T_0$. 
If the black holes have an average number density $n_\PBH{}(t)$ at time $t$ then the average number density of particles of species $l$ at the time of black hole evaporation is given by 
\bes{\label{eq:nl_tev}
    n_l(t_\mathrm{ev}) = N_l \, n_\PBH{}(t_\mathrm{ev}) 
    \;,
}
where $n_\PBH{}(t)$ is given by \eref{eq:nPBH}.  

After PBH evaporation has ended, we assume that the comoving number density of particles of species $l$ is conserved.  
In other words, particle-number-changing reactions, such as inelastic scattering and decay, are assumed to be out of equilibrium.  
This assumption implies that the relation 
\bes{\label{eq:ni_of_t}
    n_l(t) = n_l(t_\mathrm{ev}) \biggl( \frac{a(t)}{a(t_\mathrm{ev})} \biggr)^{\!\!-3} 
    \;,
}
holds, where $n_l(t)$ is the cosmological number density of particles of species $l$ at time $t$. 
The scaling $n_l \propto a^{-3}$ will hold for times $t > t_\mathrm{ev}$ and $t \lesssim \Gamma_l^{-1}$ where $\Gamma_l$ is the decay rate of particles of species $l$. 

Many of the particles created during PBH evaporation will be relativistic at the time of their production. 
For this estimate, we approximate the physical momentum of particles of species $l$ at the time of black hole evaporation as $p(t_\mathrm{ev}) \approx 2 T_0$ where $T_0$ is the initial black hole temperature.  
As the universe expands, these particles lose momentum due to cosmological redshifting such that the momentum at a later time $t > t_\mathrm{ev}$ is 
\bes{
    p_l(t) = p_l(t_\mathrm{ev}) \biggl( \frac{a(t)}{a(t_\mathrm{ev})} \biggr)^{\!\!-1} 
    \qquad \text{where} \qquad 
    p_l(t_\mathrm{ev}) \approx 2 T_0 
    \;.
}
We assume that the particles are sufficiently long lived such that they have time to become non-relativistic before they decay.  
Particles of species $l$ with mass $m_l$ become non-relativistic at time $t_{\mathrm{nr},l}$ such that $p_l(t_{\mathrm{nr},l}) = m_l$.  
In a radiation-dominated universe $a(t) \propto t^{1/2}$ and this condition resolves to 
\bes{\label{eq:tnrl}
    t_{\mathrm{nr},l} = \biggl( \frac{p_l(t_\mathrm{ev})}{m_l} \biggr)^{\!\!-2} t_\mathrm{ev}  
    \;,
}
assuming that $p_l(t_\mathrm{ev}) > m_l$ so $t_{\mathrm{nr},l} > t_\mathrm{ev}$. 
If all particle species are produced with the same physical momentum $p_l(t_\mathrm{ev}) \approx T_0$ independent of $l$, then the heavier species (larger $m_l$) will become non-relativistic earlier (smaller $t_{\mathrm{nr},l}$).  
After the particles of species $l$ have become non-relativistic ($t > t_{\mathrm{nr},l}$), we can write 
\bes{
    \rho_l(t) = m_l \, n_l(t) 
    \;,
}
where $\rho_l(t)$ is the energy density of particles of species $l$ at time $t$.  
In the next section we track the full momentum distribution by numerically solving Boltzmann's equation.  

If they are sufficiently long lived, then eventually these non-relativistic particles dominate over radiation in the cosmological energy budget, and a phase of matter domination begins. 
This is because $\rho_l(t) \propto a(t)^{-3}$ whereas $\rho_\mathrm{rad}(t) \propto a(t)^{-4}$. 
This happens at a time $t_M$ such that $\sum_l m_l n_l(t_M) = \rho_\mathrm{rad}(t_M)$. 
If the sum is dominated by the heaviest species $l = N$, then the equality resolves to 
\bes{
    t_M & = \biggl( \frac{\rho_\mathrm{rad}(t_\mathrm{ev})}{m_N n_N(t_\mathrm{ev})} \biggr)^{\!\!2} \, t_\mathrm{ev} 
    \;. 
}
Recall that $n_l(t_\mathrm{ev})$ is given by \eref{eq:nl_tev}. 

Under these assumptions, the cosmological energy fraction of particles of species $l$ at time $t$ is 
\bes{\label{eq:Omegal}
    \Omega_l(t) 
    = \frac{\rho_l(t)}{\rho_\mathrm{rad}(t) + \sum_{l=1}^{N} \rho_l(t)} 
    \;.
}
At sufficiently late times such that all species of particles in the tower are non-relativistic ($t > t_{\mathrm{nr},l}$ for all $l$) and the tower dominates over the radiation ($t > t_M$), we can write 
\bes{
    \Omega_l(t) 
    = \frac{m_l N_l}{\sum_{l=1}^{N} m_l N_l} 
    \quad (t > t_{\mathrm{nr},l}, t_M)\;.
}
Note that the dependencies on time $t$ and PBH density $n_\PBH{}$ have canceled in the ratio.  
Using the expression for $N_l$ from \eref{eq:Nl_cases} further gives 
\bes{\label{eq:Omegal_approx}
    \Omega_l = \begin{cases}
    \frac{g_l m_l}{\sum_{l^\prime=1}^{N} g_{l^\prime} m_{l^\prime}} 
    & , \quad m_1 < m_{N} < T_0 \quad \text{(whole tower always light)} \\ 
    \frac{g_l/m_l}{\sum_{l^\prime=1}^{N} g_{l^\prime}/m_{l^\prime}} 
    & , \quad T_0 < m_1 < m_{N} \quad \text{(whole tower initially heavy)}
    \end{cases} 
    \;.
} 
These expressions reveal that $\Omega_l$ takes the form of a broken power law as a function of $m_l$.  
For small masses it rises like $\Omega_l \propto m_l$, and for large masses it falls like $\Omega_l \propto m_l^{-1}$.  
In terms of the conventional stasis parametrization in \eref{eq:Omegal_stasis}, this scaling corresponds to either $\alpha = 1$ or $-1$. 
In the next sections we review the Boltzmann equation formalism and calculate the spectrum numerically to validate these approximations.\footnote{If the tower of particles responsible for stasis consists of Kaluza-Klein excitations, the inclusion of extra dimensions modifies the spectrum of the emitted particles by PBHs~\cite{Johnson:2020tiw}. These effects would become especially important if the black hole radius is comparable to the size of the compactified dimension, when the black hole ``sees" the extra dimension. }

\subsection*{Boltzmann equation}
\label{sub:boltzmann}

Let $\dd N_l = f_l(\xvec,\pvec,t) \, \dd^3 x \, \dd^3 p / (2\pi)^3$ be the average number of particles of species $l$ with spatial coordinate between $\xvec$ and $\xvec + \dd \xvec$ and with 3-momentum between $\pvec$ and $\pvec + \dd \pvec$ at time $t$.  
We call $f_l(\xvec,\pvec,t)$ the one-particle phase space distribution function.  
If the system is statistically homogeneous then $f_l$ is independent of $\xvec$.  
If the system is statistically isotropic then $f_l$ is independent of the orientation of $\pvec$ and only depends upon $p \equiv |\pvec|$. 
Under these assumptions we can write $f_l(\xvec,\pvec,t) = f_l(p,t)$, and 
\begin{equation}
    \dd n_l = \biggl( f_l(p,t) \frac{p^2}{2\pi^2} \biggr) \, \dd p \;,
\end{equation}
gives the number density of particles of species $l$ with momentum between $p$ and $p + \dd p$ at time $t$.  

The phase space distribution function $f_l$ evolves according to Boltzmann's equation, 
\begin{equation}
    L[f_l] = C[f_1, f_2, \cdots] 
    \;,
\end{equation}
where $L[f_l]$ is called the Liouville operator and $C[f_1, f_2, \cdots]$ is called the collision operator.  
In an FRW spacetime the Liouville operator takes the form~\cite{Kolb:1981hk}
\begin{equation}
    L[f_l] = \frac{\partial}{\partial t} f_l(p,t) - H(t) \, p \, \frac{\partial}{\partial p} f_l(p,t) 
    \;.
\end{equation}  
In terms of the number density per momentum interval, the Liouville operator is expressed as 
\begin{equation}
    \frac{p^2}{2 \pi^2} \times L = 
    \frac{\partial}{\partial t} \biggl( f_l(p,t) \frac{p^2}{2\pi^2} \biggr) 
    + 3 H(t) \, \biggl( f_l(p,t) \frac{p^2}{2\pi^2} \biggr) 
    - H(t) \, \frac{\partial}{\partial p} \biggl( p f_l(p,t) \frac{p^2}{2\pi^2} \biggr) 
    \;.
\end{equation}
The Boltzmann equation equates the Liouville operator with the collision operator associated with any processes that change particle number or shuffle particles (elastically) between momentum bins.  
The production of particles via PBH evaporation corresponds to a source term 
\begin{equation}
    \frac{p^2}{2 \pi^2} \times C 
    = \frac{g_l}{8 \pi^2} \, \frac{p^3}{E_l(p)} \, \frac{1}{\ee^{E_l(p) / T_\PBH{}} + s_l} \, 4 \pi b_S^2 \, n_\PBH{} 
    \;,
\end{equation}
which is independent of the $f_l$.  
We neglect the scattering and decay channels, which would correspond to additional terms in $C[f_1, f_2, \cdots]$.  
Without loss of generality, we can write 
\bes{
    f_l(p,t) = F_l(p_c, t) \Bigr|_{p_c = a(t) p} 
    \;,
}
where the first argument of $F_l$ is the comoving momentum of a particle that has momentum $p$ at time $t$. 
Putting together these pieces yields Boltzmann's equation 
\bes{\label{eq:Boltzmann_eqn}
    & \frac{\partial}{\partial t} F_l(p_c, t) 
    = S_l(p_c, t) 
    \quad \text{where} \quad 
    S_l(p_c, t) 
    = g_l \, \frac{p_c}{\sqrt{p_c^2 + a(t)^2 m_l^2}} \, \frac{\pi b_S(t)^2 \, n_\PBH{}(t)}{\ee^{\sqrt{p_c^2 + a(t)^2 m_l^2} / a(t) T_\PBH{}(t)} + s_l} 
    \;,
}
where the partial time derivative only acts on the second argument of $F_l$.  
In the next section we solve this equation numerically to calculate the development of the emission spectrum while the black holes are evaporating.  

\subsection*{Development of the emission spectrum}
\label{sub:spectrum}

We numerically solve Boltzmann's equation \eqref{eq:Boltzmann_eqn} along with the initial condition $F_l(p_c,t_0) = 0$ for all $p_c \geq 0$, which expresses the assumption that no particles are produced before the PBHs are formed (and begin evaporating) at time $t_0$. 
In order to understand the results, we find is useful to consider two cases corresponding to particles that are ``always light'' ($m_l < T_0$) and particles that are ``initially heavy'' ($m_l > T_0$); see also \eref{eq:Nl_cases}.  
To understand how the spectrum develops over the black hole's lifetime, we first discuss the source term $S_l(p_c,t)$, and then present the numerical results. 

\subsubsection*{Source term}

The source term $S_l(p_c,t)$ is composed of four factors: 
\bes{\label{eq:source_factors}
    S_l(p_c,t)
    \propto \undert{ \frac{p_c}{\sqrt{p_c^2 + a(t)^2 m_l^2}}}{\text{factor~\#1}} \quad \undert{ \bigl( \ee^{\sqrt{p_c^2 + a(t)^2 m_l^2} / a(t) T_\PBH{}(t)} + s_l \bigr)^{-1}}{\text{factor~\#3}}  \,\undert{ \quad b_S(t)^2 \quad}{\text{factor~\#4}} \,\undert{ n_\PBH{}(t) \quad}{\text{factor~\#2}} 
    \;.
}
\begin{itemize}
\item \textbf{factor~\#1: speed.}  This factor is the speed of a particle with mass $m_l$ and comoving momentum $p_c$ at time $t$.  For non-relativistic particles ($p_c \ll a(t)m_l$) this factor scales linearly with momentum $\propto p_c^1$, and decreases with time $\propto a(t)^{-1}$ as the scale factor grows.  For relativistic particles ($p_c \gg a(t)m$), this factor is approximately independent of $p_c$ and constant in time.
\item \textbf{factor~\#2: number density.}  This factor is the number density of PBHs.  It decreases with time $\propto a(t)^{-3}$ as the universe expands and the scale factor grows; see \eref{eq:nPBH}. 
    \item \textbf{factor~\#3: thermal factor.}  This factor is the Bose-Einstein or Fermi-Dirac thermal distribution function.  It is responsible for the exponential drop in $F(p_c,t)$ toward large $p_c$.  We identify $p_c = p_\mathrm{exp}(t)$ to be the comoving momentum where the kinetic energy equals the thermal energy: 
    \begin{align}\label{eq:pexp}
        \frac{\sqrt{p_\mathrm{exp}(t)^2 + a(t)^2 m_l^2}}{a(t)} - m_l = 2 T_\PBH{}(t) 
        \quad \Rightarrow \quad 
        p_\mathrm{exp}(t) = a(t) \, \sqrt{\bigl( 2 T_\PBH{}(t) + m_l \bigr)^2 - m_l^2}
        \;.
    \end{align}
    For a particle that is light ($m_l < T_\PBH{}$) this is approximately $p_\mathrm{exp}(t) \approx 2 a(t) T_\PBH{}(t)$, and for a particle that is heavy ($m_l > T_\PBH{}$) this is approximately $p_\mathrm{exp}(t) \approx 2 a(t) \sqrt{m_l T_\PBH{}(t)}$.  Also recall how particles that are initially heavy ($m_l > T_0$) will eventually become light at a time $t_m = t_0 + (1 - T_0^3 / m_l^3) \, \tau_\PBH{}$, given by \eref{eq:tm}.  

\item \textbf{factor~\#4: Schwarzschild radius.}  This factor is proportional to the black hole's Schwartzchild radius $b_S(t)^2 = \tfrac{27}{4} R_S(t)^2$, given by \eref{eq:RS}.  As $t$ approaches the black hole's evaporation time $t_\mathrm{ev}$, the Schwartzchild radius decreases toward zero, suppressing subsequent particle emission.  
\end{itemize}

\subsubsection*{Case 1: always light ($m_l < T_0$)}

\begin{figure}[t]
    \centering
    \includegraphics[width=0.7\linewidth]{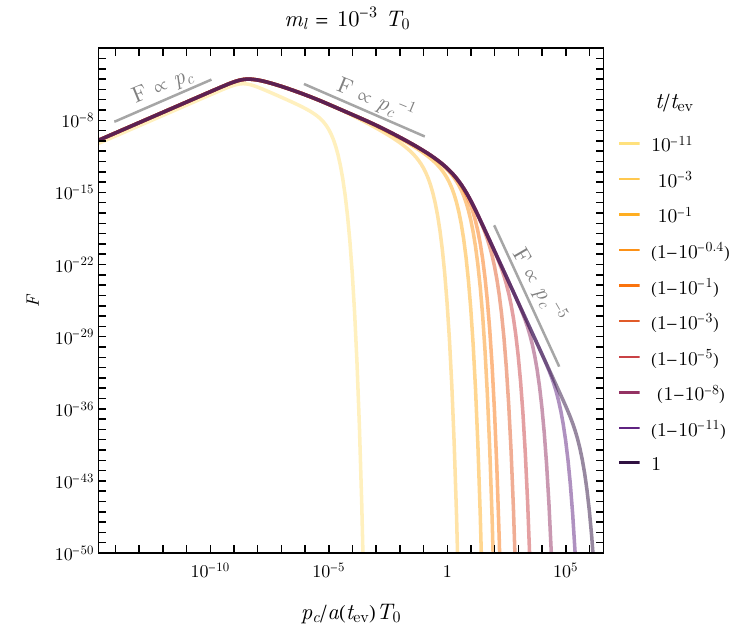}
    \caption{\label{fig:lightcase}
    Evolution of the distribution function $F(p_c,t)$ for a particle species that is always light $m_l < T_0$.  The spectrum is a function of the comoving momentum $p_c$, which is expressed in units of $a(t_\mathrm{ev}) T_0$ where $a(t_\mathrm{ev})$ is the FRW scale factor at the time when the black hole evaporates and $T_0$ is the initial black hole temperature.  The various colored curves show $F(p_c,t)$ at different times $t$ between black hole formation $t = t_0$ and black hole evaporation $t = t_\mathrm{ev} \lesssim t_0 + \tau_\PBH{}$.  
    For this numerical example, the parameters are chosen to be $M_0 = 1$ gram, $\beta = 10^{-8}$, $g_\ast = g_\SM{} + 100 = 206.75$, $m_l = 10^{-3} T_0$, $s_l = -1$, and $g_l = 1$. 
    }
\end{figure}

For the case in which a black hole with initial mass $M_0$ and temperature $T_0$ emits a particle of mass $m_l$ that is always light $m_l < T_0$, \fref{fig:lightcase} shows the phase space distribution function $F(p_c,t)$ as a function of comoving momentum $p_c$ at several different times $t$.  
During the first few time steps, the distribution function scales as $F \propto p_c^1$ for small comoving momentum, $\propto p_c^{-1}$ for intermediate momentum, and $\propto \mathrm{exp}(-p_c)$ for large momentum.  
At small momentum $p_c < a(t) m_l$, the rising behavior $F \propto p_c^1$ follows from factor~\#1 in the source \eqref{eq:source_factors}.  
Since factor~\#2, which goes as $n_\PBH{}(t) \propto a(t)^{-3}$, is monotonically decreasing with time, this part of the spectrum receives its largest contribution at the earliest time, and afterwards it is approximately static. 
At large momentum $p_c > p_\mathrm{exp}(t) \approx 2 a(t) T_\PBH{}(t)$, the falling behavior $F \propto \mathrm{exp}(-p_c / a(t) T_\PBH{}(t))$ follows from factor~\#3.  
Since $a(t) T_\PBH{}(t)$ increases with time, this break in the spectrum evolves toward larger $p_c$. 
At intermediate momentum $a(t) m_l < p_c < a(t) T_\PBH{}(t)$ where factor~\#1 is approximately constant, the falling behavior $F \propto p_c^{-1}$ follows from factor~\#3, which goes as $(\ee^{p_c / a T_\PBH{}} -1)^{-1} \approx (p_c / a T_\PBH{})^{-1}$ for bosons ($s_l = -1$).  
At later times the distribution function develops an additional scaling $F \propto p_c^{-5}$. 
This is a consequence of factor~\#4, which goes as $b_S(t)^2 \propto R_S(t)^2$, and which begins to decrease rapidly as $t$ nears the evaporation time $t_\mathrm{ev}$.  
The location of the break between the $p_c^{-1}$ and $p_c^{-5}$ branches is important, because it is where the spectrum $\propto p_c^3 F$ is peaked.  
We use $p_{\ast,l} = p_c|_\text{at break}$ to denote this scale; see \eref{eq:pastl}. 
It is approximately given by $p_{\ast,l} \approx p_\mathrm{exp}(t_2)$ at the time $t_2$ when the Schwartzchild radius has decreased by a factor of $2$.  
Recall that $p_\mathrm{exp}(t) \approx 2 a(t) T_\PBH{}(t)$ for $m_l < T_\PBH{}(t)$, and we can write $p_\mathrm{exp}(t_2) \approx 2 a(t_\mathrm{ev}) T_0$, since the black hole temperature doesn't grow much between $t_0$ and $t_2$, and since the scale factor doesn't grow much between $t_2$ and $t_\mathrm{ev}$.  

\subsubsection*{Case 2:  initially heavy ($m_l > T_0$) }

\begin{figure}[t]
    \centering
    \includegraphics[width=0.7\linewidth]{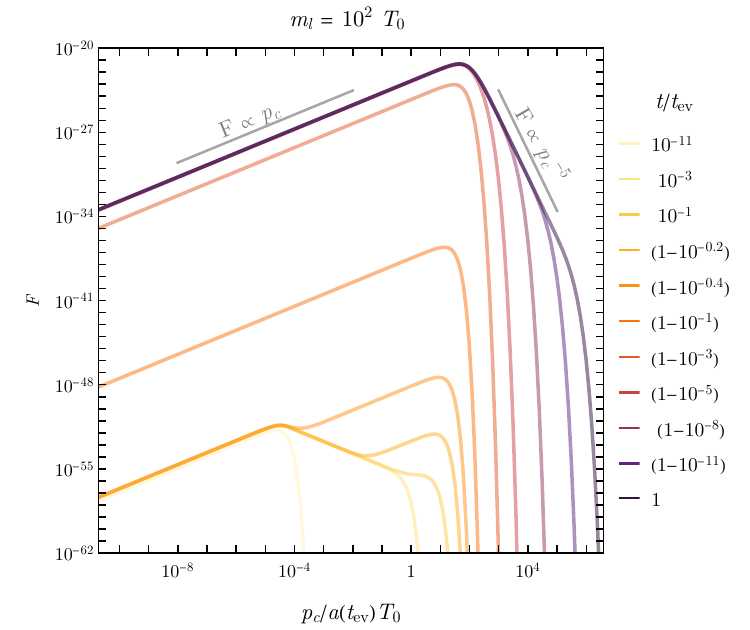}
    \caption{\label{fig:heavycase}
    Evolution of the distribution function $F(p_c,t)$ for a particle species that is initially heavy $m_l > T_0$.  
    Same as \fref{fig:lightcase} except $m_l = 10^{2} T_0$. 
 } 
\end{figure}

For the case in which a black hole emits a particle that is initially heavy $m_l > T_0$, \fref{fig:heavycase} shows the phase space distribution function $F(p_c,t)$.  
For the first three time steps, $F$ displays the same $p_c^1$, $p_c^{-1}$, and $\mathrm{exp}(-p_c)$ scalings that also appear on \fref{fig:lightcase}. 
However, the overall size of $F$ is much smaller, because factor~\#3 is on the order of $\mathrm{exp}(-m_l / a T_\PBH{}) \ll 1$.    
During these time steps, the distribution function is peaked at $p_c = p_\mathrm{exp}(t_0)$, see \eref{eq:pexp}, which marks the momentum scale where the kinetic energy of the particle becomes comparable to the initial black hole temperature.   
As time goes on, the black hole temperature increases, allowing for the production of particles with higher momentum, and so the spectrum extends towards larger $p_c$ up to $p_\mathrm{exp}(t)$.    

For the fourth through eighth time steps, $F(p_c,t)$ grows rapidly across a wide range of momentum.  
This is because the black hole temperature $T_\PBH{}(t)$ has grown larger than $m_l$, and the exponential Boltzmann suppression in factor~\#3 is removed.    
After the eighth time step, which corresponds to the time scale $t_m$ from \eref{eq:tm}, the spectrum stops rising at small $p_c$. 
For the remaining time steps, the spectrum develops a $p_c^{-5}$ power law behavior, just as in \fref{fig:lightcase}.  
The break between the $p_c^1$ and $p_c^{-5}$ branches occurs when $p_{\ast,l} = p_c|_\text{at break} \approx p_\mathrm{exp}(t_m) = 2 a(t_m) m_l$; see \eref{eq:pastl}.  
Here we used \eref{eq:pexp} and $T_\PBH{}(t_m) = m_l$, which defines $t_m$.  
To summarize, the black hole is an inefficient radiator until $T_\PBH{}(t) > m_l$, and the spectrum that arises subsequently is approximately the same as a black hole that forms with $T_0 = m_l$. 

\subsection*{Spectrum and typical momentum}
\label{sub:spectrum}

\begin{figure}[t]
    \centering
    \includegraphics[width=0.49\linewidth]{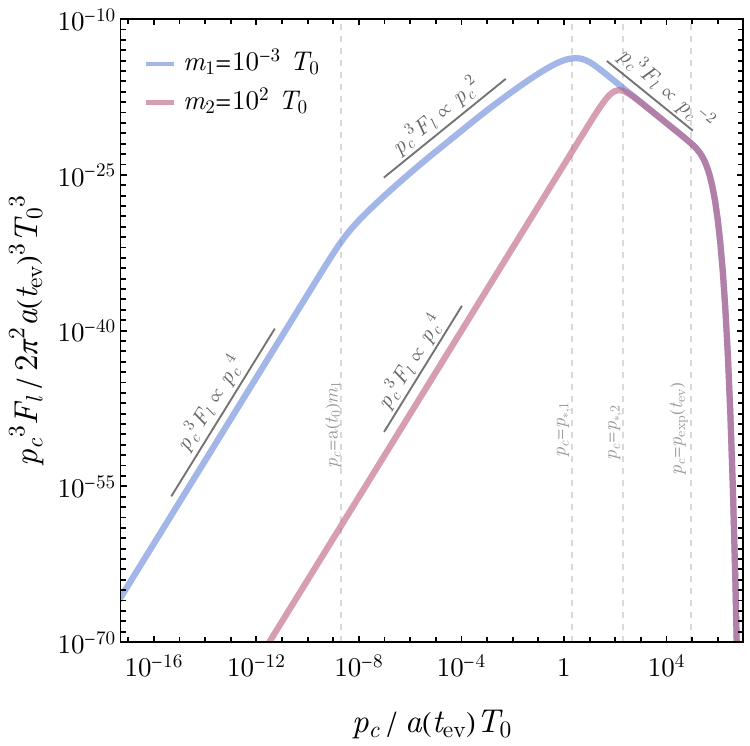}
    \hfill
    \includegraphics[width=0.49\linewidth]{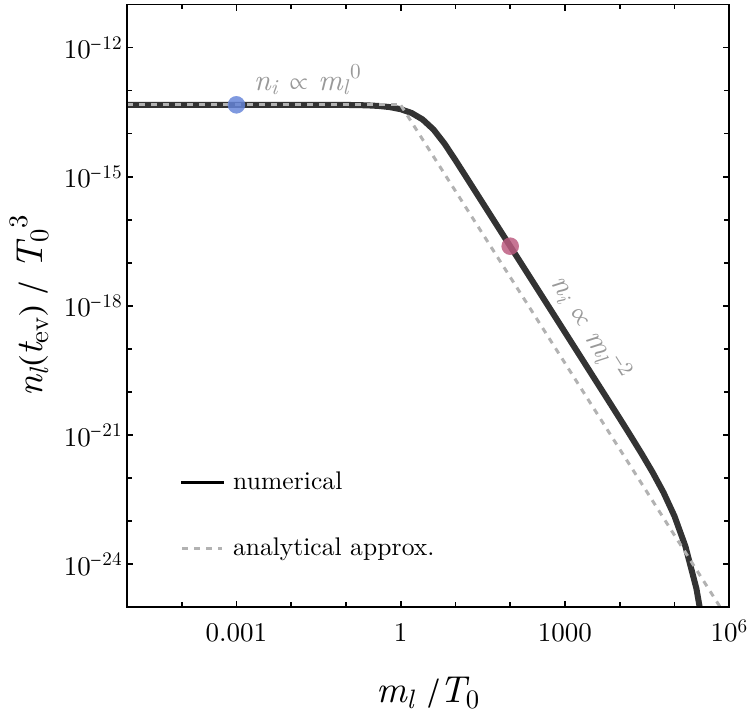}
    \caption{\label{fig:Fplot}\label{fig:both_time}
    \textit{Left:} 
    Spectrum of emitted particles of species $l$ with mass $m_l$ when a population of black holes having initial mass $M_0$, initial temperature $T_0$, and initial density $n_{\PBH{},0}$ are completely evaporated at time $t_\mathrm{ev}$.  The product $\tfrac{p_c^3}{2\pi^2} F_l(p_c, t_\mathrm{ev})$ represents the number density at time $t_\mathrm{ev}$ per log-spaced comoving momentum interval. Parameters are the same as \fref{fig:lightcase} except $m_l = m_1 = 10^{-3} T_0$ (blue) and $m_l = m_2 = 10^{2} T_0$ (red).  
    \textit{Right:} Number density of emitted particles of species $l$ with mass $m_l$.  Besides $m_l$ the other parameters are identical to the left panel.  The dashed lines plot the analytical approximation in \eref{eq:nl_tev_cases}. The blue ($m_l=10^{-3}T_0$) and red ($m_l=10^{2}T_0$) dots highlight the masses shown in the left panel. 
    }
\end{figure}

The spectrum of particles emitted by the population of PBHs is calculated by numerically integrating Boltzmann's equation \eqref{eq:Boltzmann_eqn} from time $t = t_0$ when the black holes form until time $t = t_\mathrm{ev}$ when the black holes have completely evaporated. 
In the left panel of \fref{fig:Fplot} we show the results of this calculation for two representative parameter points.   
We show two spectra corresponding to different choices of the emitted particle's mass:  for $m_l = 10^{-3} T_0$ the emitted particle is always light (blue), and for $m_l = 10^2 T_0$ the emitted particle is initially heavy (red).  
Observe that the spectrum is much smaller for the heavier particle, since its production is Boltzmann suppressed until the final stages of black hole evaporation when $T_\PBH{}(t)$ rises above $m_l$. 
The spectrum $p_c^3 F$ finds a maximum at the scale 
\begin{align}\label{eq:pastl}
    p_{c\ast,l}  
    & \approx \begin{cases} 
    2 a(t_\mathrm{ev}) T_0 
    & , \quad m_l < T_0 \quad \text{(always light)} \\ 
    2 a(t_m) m_l  
    & , \quad m_l > T_0 \quad \text{(initially heavy)} 
    \end{cases} 
    \;, 
\end{align}
where $t_m = t_0 + (1 - T_0^3 / m_l^3) \, \tau_\PBH{}$ is the time when $T_\PBH{}(t_m) = m_l$, as in \eref{eq:tm}.    
Toward large values of $p_c$ the spectrum falls exponentially; this is because we stop the evolution at time $t = t_\mathrm{ev}$ when the black hole mass is $M_\PBH{}(t_\mathrm{ev}) = \MPl$ and the black hole temperature is $T_\PBH{}(t_\mathrm{ev}) = \MPl/8\pi$, so the emission of particles with $p_c/a(t_\mathrm{ev}) \gtrsim 2 T_\PBH{}(t_\mathrm{ev}) \approx 800 T_0$ remains Boltzmann suppressed. 

In the right panel of \fref{fig:Fplot} we show the number density $n_l(t_\mathrm{ev})$ of particles of species $l$ with mass $m_l$ that are emitted from the population of PBHs.  
For particles that are always light ($m_l < T_0$) their density is insensitive to their mass, but instead governed by the initial black hole mass $M_0$, temperature $T_0$, and parameter $\beta$ that sets the initial density $n_{\PBH{},0}$.  
For particles that are initially heavy ($m_l > T_0$) their density decreases with increasing mass, since these particles are only produced efficiently near the end of the black hole lifetime when $T_\PBH{}(t)$ rises to overtake the mass.  
In both cases, the numerical results agree well with the analytical approximation in \erefs{eq:Nl_cases}{eq:nl_tev}, 
\begin{align}\label{eq:nl_tev_cases}
    n_l(t_\mathrm{ev}) = \begin{cases}
    \frac{15 \zeta(3)}{\pi^4} \frac{g_l}{g_\ast} \frac{M_0}{T_0} n_\PBH{}(t_\mathrm{ev})
    & , \quad m_l < T_0 \quad \text{(always light)} \\ 
    \frac{15 \zeta(3)}{\pi^4} \frac{g_l}{g_\ast} \frac{M_0 T_0}{m_l^2} n_\PBH{}(t_\mathrm{ev}) 
    & , \quad m_l > T_0 \quad \text{(initially heavy)}
    \end{cases}
    \;,
\end{align}
which are shown as dashed-gray lines. 

When the black holes have fully evaporated, it is interesting to ask whether the particles they emitted are relativistic or non-relativistic.  
To answer this question we consider two measures of the particles' momentum.  
First we consider the scale $p_c$ at which the spectrum $\propto p_c^3 F_l(p_c, t_\mathrm{ev})$ reaches its maximum, and we evaluate the corresponding physical momentum $p_{\ast,l} = p_c / a(t_\mathrm{ev})$ given by \eref{eq:pastl}, which represents the momentum of the most abundant particles in the system.  
Second we integrate over the full phase space to calculate the root-mean-square momentum
\begin{equation}\label{eq:prms}
    p_{\mathrm{rms},l} = \biggl[ \frac{1}{N_l} \int_0^\infty \! \frac{\dd p_c}{p_c} \left(\frac{p_c}{a(t_\mathrm{ev})}\right)^2 \frac{p_c^3}{2\pi^2} F_l(p_c,t_\mathrm{ev}) \biggr]^{1/2} 
    \;,
\end{equation}
where $N_l = \int_0^\infty \! \tfrac{\dd p_c}{p_c} \, \frac{p_c^3}{2\pi^2} F_l(p_c,t_\mathrm{ev})$ is the number of particles of species $l$ emitted from the black hole.   

\begin{figure}[t]
    \centering
    \includegraphics[width=0.49\linewidth]{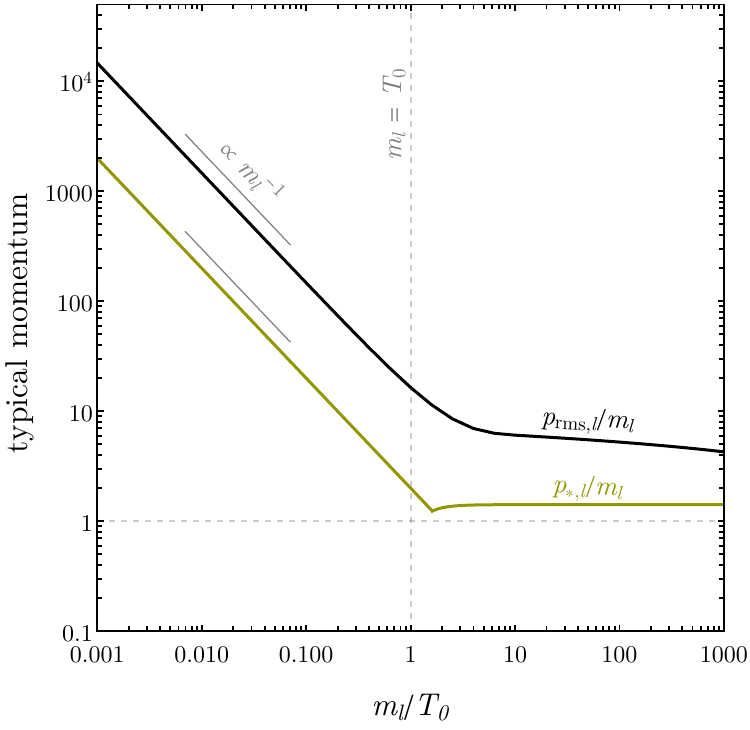}
    \caption{\label{fig:momentum}
    Typical momentum of particles of species $l$ with mass $m_l$ at time $t_\mathrm{ev}$ when the black holes evaporate.  We show two different measures of the typical momentum.  The green curve shows $p_{\ast,l}$, the physical momentum corresponding to the peak of the spectrum $\propto p_c^3 F_l(p_c,t_\mathrm{ev})$.  The black curve shows $p_{\mathrm{rms},l}$, the root-mean-square momentum calculated by integrating the full spectrum. The vertical dashed lines indicate $m_l = T_\PBH{}(t_0) = T_0$ the initial black hole temperature at formation. Besides $m_l$ the other parameters are identical to \fref{fig:Fplot}.
   }
\end{figure}

In \fref{fig:momentum} we show $p_{\ast,l}/m_l$ and $p_{\mathrm{rms},l}/m_l$, which can be compared with $1$ to assess whether particles of species $l$ with mass $m_l$ are produced with typical velocities that are relativistic or non-relativistic.  
Particles that are always light ($m_l < T_0$) are relativistic at the time $t_\mathrm{ev}$ when the black holes evaporate, and typically $p_{\ast,l} \approx 2T_0$ and $p_{\mathrm{rms},l} \approx 15 T_0$. 
The larger value of $p_{\mathrm{rms},l}$ is due to the log-enhancement from the $p_c^3 F_l \propto p_c^{-2}$ branch of the spectrum. 
Particles that are initially heavy ($m_l > T_0$) are only marginally relativistic at the time of evaporation: we find $p_{\ast,l} \approx 1.4 m_l$ and $p_{\mathrm{rms},l} \approx 5.4 m_l$ for the two momenta.  
The more relativistic particles will require a longer period of cosmological expansion to make them non-relativistic; see \eref{eq:tnrl}. 

\subsection*{Resultant abundance of emitted particles}
\label{sub:abundance}

\begin{figure}[t]
    \centering
    \includegraphics[width=0.49\linewidth]{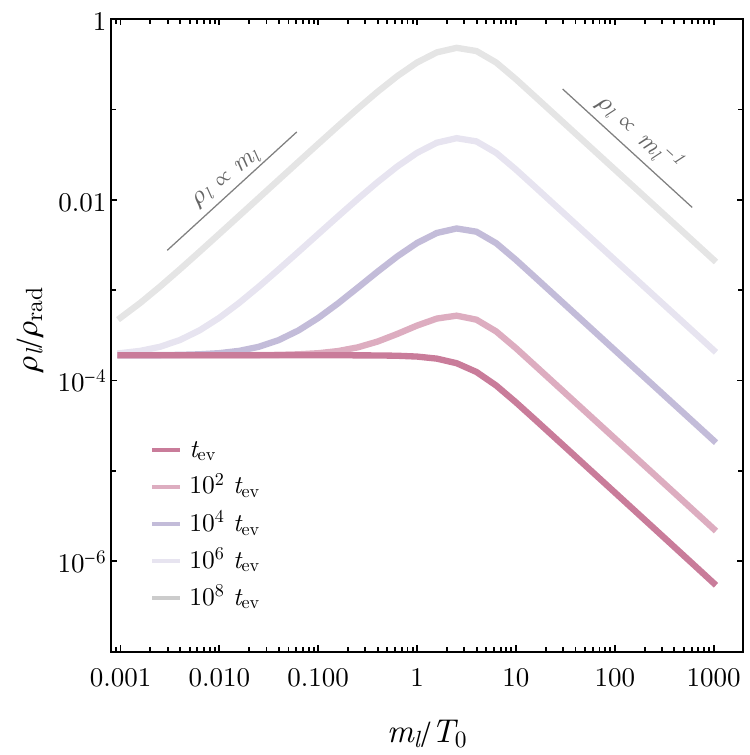} \hfill
    \caption{\label{fig:Omega}
    Energy fraction $\rho_l(t)/\rho_\mathrm{rad}(t)$ for particles of species $l$ at several different times $t$.  Besides $m_l$ the other parameters are identical to \fref{fig:lightcase} 
    }
\end{figure}

Finally we calculate the cosmological energy fraction $\Omega_l(t)$ carried by particles of species $l$ at time $t$.  
To do so, we evaluate $\Omega_l(t)$ using \eref{eq:Omegal}: 
\bes{
    \Omega_l(t) = \frac{\rho_l(t)}{\rho_\mathrm{rad}(t) + \rho_\mathrm{tower}(t)} 
    \qquad \text{where} \qquad 
    \rho_\mathrm{tower}(t) = \sum_{l=1}^{N} \rho_l(t)
    \;,
    \nonumber
}
where $\rho_\mathrm{rad}(t)$ is the radiation energy density \eqref{eq:rho_rad}, where $\rho_l(t)$ is calculated by integrating the spectrum 
\begin{equation}
    \rho_l(t) = a(t)^{-3} \int_0^\infty \! \frac{\dd p_c}{p_c} \, E_l(p_c,t) \, \frac{p_c^3}{2\pi^2} F_l(p_c,t_\mathrm{ev}) 
    \qquad (t \geq t_\mathrm{ev}) 
    \;,
\end{equation}
and where the energy is $E_l(p_c,t) = \sqrt{p_c^2/a(t)^2+m_l^2}$ at time $t$. 

In \fref{fig:Omega} we show how the ratio $\rho_l(t) / \rho_\mathrm{rad}(t)$ varies with mass $m_l$ at several different times.  
At time $t = t_\mathrm{ev}$ just after the black holes have evaporated, the emitted particles are subdominant to the radiation and $\rho_l/\rho_\mathrm{rad} \ll 1$ for all masses.  
Initially the emitted particles are mostly relativistic so $\rho_l(t) \propto a(t)^{-4}$ as well as $\rho_\mathrm{rad}(t) \propto a(t)^{-4}$, such that the ratio is static.  
As time goes by, the momentum of the emitted particles decreases due to the cosmological expansion (redshifting).  
Once the particles of species $l$ become non-relativistic at time $t_{\mathrm{nr},l}$, see \eref{eq:tnrl}, their energy density decreases only as $\rho_l(t) \propto a(t)^{-3}$, and $\rho_l(t)/\rho_\mathrm{rad}(t) \propto a(t)^1$ begins to grow.      
In the last couple time steps, the particles at all masses are non-relativistic.  

To evaluate $\Omega_l(t)$ we must specify not only the mass $m_l$ of particles of species $l$, but we must specify the full mass spectrum, which enters through the sum in $\rho_\mathrm{tower}$.  
To do this, we use parametrization in \eref{eq:m_l}, which reads $m_l = m_1 + (l-1)^\delta \, \Delta m$ for $l \in \{ 1, 2, \cdots, N \}$. 
We consider two cases:  (1) the whole tower is always light $m_1 = \Delta m = 10^{-3} T_0$ and (2) the whole tower is initially heavy $m_1 = \Delta m = 10^{2} T_0$.  
In both cases we take $\delta = 1$ and $N = 100$.  

\begin{figure}[h!]
    \centering
    \includegraphics[width=0.44\linewidth]{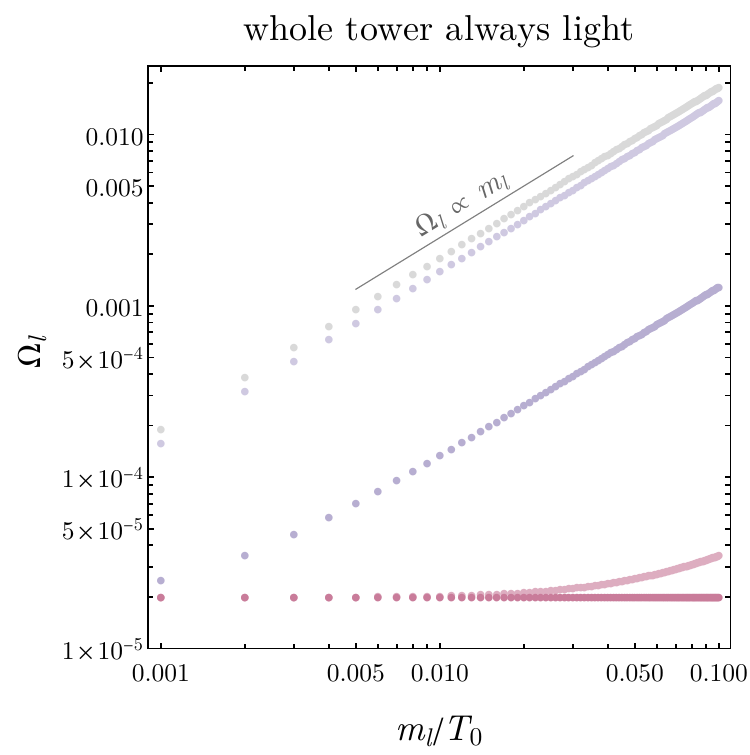}
    \hfill
    \includegraphics[width=0.54\linewidth]{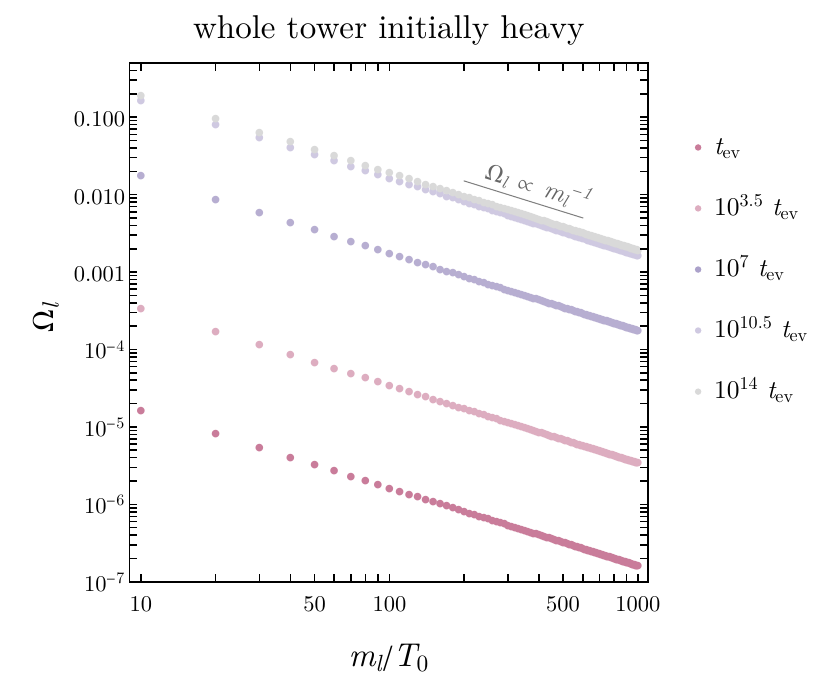}
    \caption{\label{fig:Option2} 
    Cosmological energy fraction $\Omega_l$ for particles of species $l$ with mass $m_l = m_1 + (l-1)^\delta \Delta m$ and other parameters identical to \fref{fig:lightcase}.  
    \textit{Left:}  We consider a tower of $N = 100$ ``always light'' particles with $m_1 = \Delta m = 10^{-3} T_0$, $\delta = 1$, and $g_\ast = g_\SM{} + N = 206.75$.  
    \textit{Right:}  We consider a tower of $N = 100$ ``initially heavy'' particles with $m_1 = \Delta m = 10^{2} T_0$, $\delta = 1$, and $g_\ast = g_\SM{} = 106.75$.  
    }
\end{figure}

\Fref{fig:Option2} shows the the cosmological energy fraction $\Omega_l(t)$ for these two scenarios.  
During the first few time steps $\Omega_l(t)$ grows for the non-relativistic particles since $\rho_l(t) / \rho_\mathrm{rad}(t) \propto a(t)^1$.  
Eventually at time $t_M$ the tower carries the same amount of energy as the radiation, $\rho_\mathrm{tower}(t_M) = \rho_\mathrm{rad}(t_M)$.  
Afterward $\Omega_l(t)$ stops growing since $\rho_l(t) / \rho_\mathrm{tower}(t) \propto a(t)^0$.    

To draw closer connection with the initial conditions needed for stasis, we recall the parametrization $\Omega_l = (m_l / m_1)^\alpha \Omega_1$ from \eref{eq:Omegal_stasis}.  
For particles that are always light ($m_l < T_0$), the energy fraction scales as the first power of mass: $\Omega_l \propto m_l$, which corresponds to $\alpha = 1$, and for particles that are initially heavy ($m_l > T_0$), the energy fraction scales as $\Omega_l \propto m_l^{-1}$, which corresponds to $\alpha = -1$.  
The conclusion is that populating a tower of massive particle by PBH evaporation can result in 
\begin{align}\label{eq:alpha_PBH}
    \Omega_l \propto m_l^\alpha \quad \text{with} \quad \alpha = +1 \quad \text{or} \quad \alpha = -1 
    \;.
\end{align}
It is also interesting to note that if the range of masses in the tower straddles $m_l = T_0$ then $\Omega_l$ takes the form of a broken power law with $\alpha = +1$ for small masses and $\alpha = -1$ for large masses.  
This observation motivates a study of stasis in models with non-monomial relations between $\Omega_l$ and $m_l$. 


\subsection*{Decay of the tower and starting stasis}
\label{sub:decay}

\begin{figure}[t]
    \centering
    \includegraphics[width=0.7\linewidth]{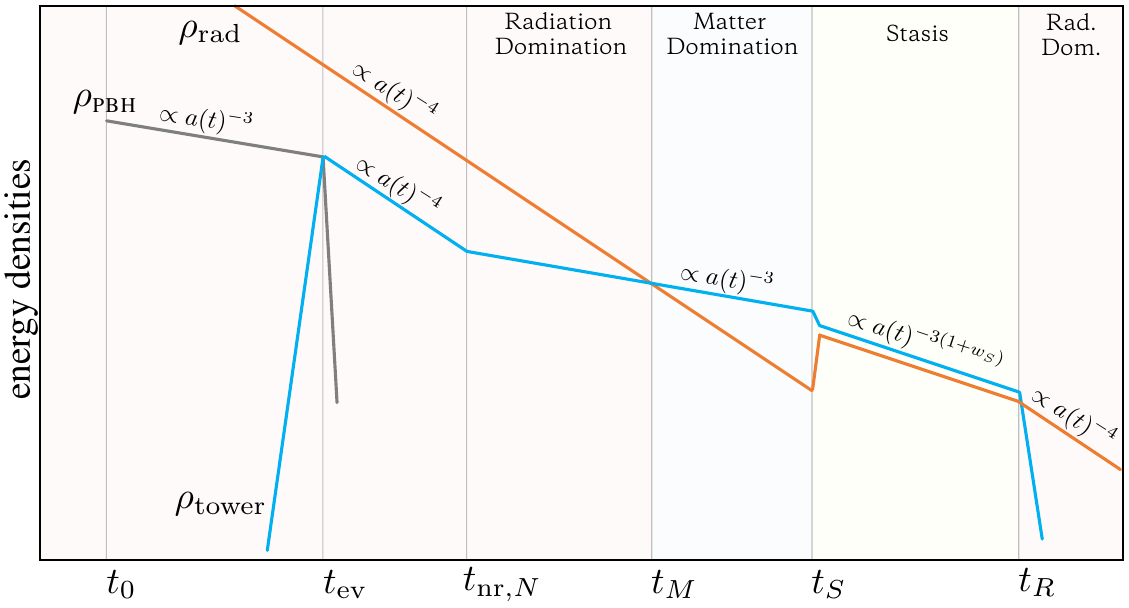}
    \caption{\label{fig:timeline}
    A schematic illustration of the evolving energy densities.  Time increases to the right on the horizontal axis, which labels: black hole formation $t_0$, black hole evaporation $t_\mathrm{ev}$, the heaviest species $l=N$ becomes non-relativistic $t_{\mathrm{nr},N}$, matter domination begins $t_M$, stasis begins $t_S$, and finally reheating begins a second phase of radiation domination $t_R$. The lines indicate the energy density carried by the radiation (orange), the PBHs (gray), and the tower of massive particles (blue).  }
\end{figure}

So far, we have discussed how a tower of stable states with different masses can be populated by a population of PBHs, along with their spectrum. To achieve a period of stasis, it is necessary for the particles in the tower  to dominate the energy density of the Universe and then decay into radiation. Therefore, we assume that the particles in the tower are unstable with decay rates given by a power-law relation (c.f. \eref{eq:Gamma_l}). In our scenario, the thermal history of the Universe, depicted in \fref{fig:timeline}, can be summarized as follows: at some point after reheating, the Universe is only filled with Standard Model radiation. At time $t=t_0$, a fraction $\beta\ll 1$ of the energy density of the Universe collapses into a population of PBHs with masses of the order of the horizon mass at that time. PBHs remain the subdominant component of the energy density of the Universe and eventually evaporate at time $t=t_{\rm ev}$ and release their energy in the form of Standard Model particles and particles in the tower. According to the results of \sref{sub:spectrum}, particles with masses below the initial temperature of the PBHs are produced relativistically. Even particles with masses above this temperature typically have momenta a few times their mass, and are therefore relativistic to some extent. Provided that the particles in the tower interact very weakly with the radiation in the Universe, they lose energy solely due to redshifting prior to their decay. Particles at level $l$ become non-relativistic at $t=t_{{\rm nr},l}$ after sufficient expansion, which is given by:
\begin{align}\label{eq:non-rel}
    \frac{a(t_{{\rm nr},l})}{a(t_{\rm ev})} \sim \begin{cases}
    T_0/m_l
    & , \quad m_l < T_0  \\ 
    \mathrm{few} 
    & , \quad m_l > T_0 
    \end{cases}
    \;.
\end{align}
Particles in the tower become non-relativistic in sequence, starting with the heavier states. If a large enough number of heavy states become non-relativistic before they decay, they may come to dominate over radiation and initiate an early matter-dominated era at time $t=t_M$. This sets the stage for stasis, which proceeds with the successive decay of heavy states into radiation. Hence, the lifetime of the heaviest state, $\tau_N = \Gamma_N^{-1}$ roughly marks the beginning of stasis, $t_S$, i.e., $t_S \sim \tau_N = \Gamma_N^{-1}$. Eventually, stasis ends when the lightest particle in the tower decays, having lifetime $\tau_1 = \Gamma_1^{-1}$, and at this point Universe becomes radiation-dominated for a second time at time $t = t_R = \tau_1 = \Gamma_1^{-1}$.
To remain consistent with the successful predictions of Big Bang Nucleosynthesis (BBN), the temperature of the Universe at $t_R$ should be above the MeV range~\cite{Hasegawa:2019jsa,Kawasaki:2000en}, and therefore $\Gamma_1/{\rm MeV}\gtrsim 10^{-22}$. 

\begin{figure}[t]
    \centering
    \includegraphics[width=0.48\linewidth]{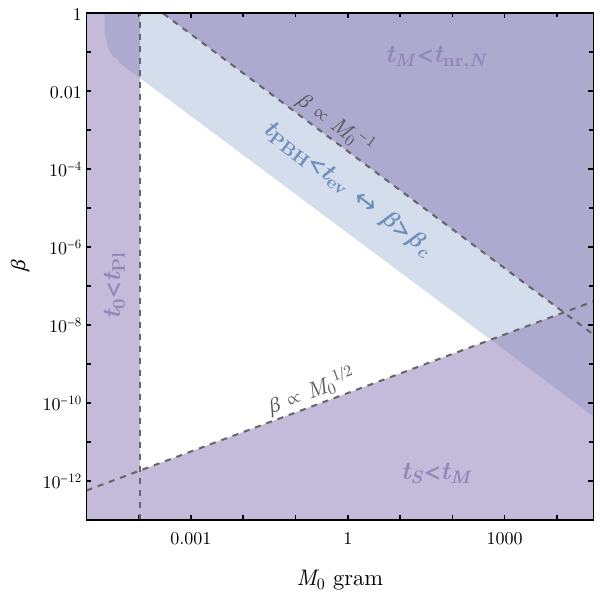}
    \hfill
    \includegraphics[width=0.48\linewidth]{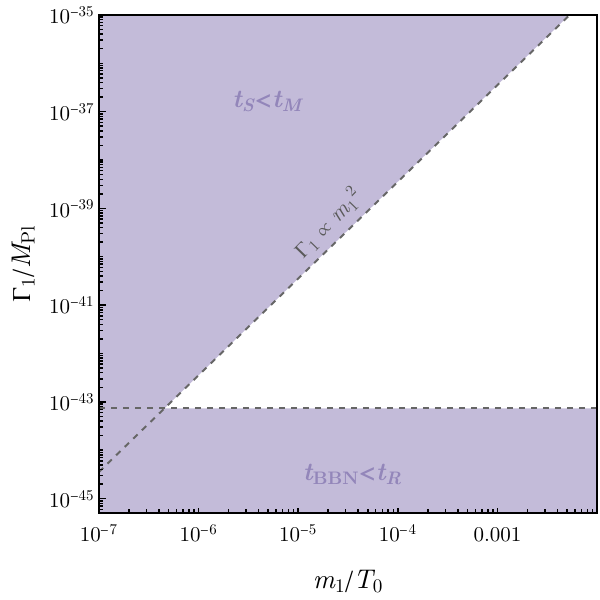}
    \caption{\label{fig:param_space} 
    The shaded regions corresponds to parameter combinations that violate the required time ordering of cosmological events. The timeline corresponds to the sequence: $t_\mathrm{Pl} < t_0 < t_\mathrm{ev} < t_{\mathrm{nr},N} < t_{M} < t_{S} < t_{R} < t_\BBN{}$ and $t_\mathrm{ev} < t_\PBH{}$ as in \eref{eq:tPBH}. \textit{Left:}  We vary $M_0$ and $\beta$ while holding fixed $\Gamma_1 = 10^{-40} \, \mathrm{GeV}$, $m_1 = \Delta m = 10^{-3} T_0$, $N = 100$, $m_N = 10^{-1} T_0$, and $\delta = 1$.  \textit{Right:}  We vary $\Gamma_1$ and $m_1 = \Delta m$ while holding fixed $M_0 = 1 \, \mathrm{gram}$, $\beta = 10^{-8}$, $\delta = 1$, $N = 100$, and $m_N = 100 m_1$.  The other parameters same as in \fref{fig:lightcase}.
    }
\end{figure}

\Fref{fig:param_space} shows the viable regions of parameter space that are consistent with the cosmological time ordering illustrated in \fref{fig:timeline}. 
In the left panel, we vary $M_0$ and $\beta$ which are parameters associated with the primordial black holes. 
The region labeled $t_0 < t_\mathrm{Pl}$ is inconsistent, because PBHs would form before the Planck time, where our classical treatment of spacetime is no longer valid. 
The region labeled $t_S < t_M$ is inconsistent, because for small values of $\beta$ there is a low PBH energy density and, consequently, a suppressed number density of particles. 
As a result, the produced particles would take too long to dominate over radiation and would decay before achieving domination. 
This effect increases for larger masses, because the number density of PBHs falls linearly with $M_0$.  
The region labeled $t_M < t_{\mathrm{nr},N}$ is inconsistent, because the large amount of produced particles would dominate the energy density before becoming non-relativistic, contradicting our assumption that they behave as dust during their domination era. 
This occurs at large $\beta$, where the PBH abundance, and thus the particle production, is high. 
Finally, the region labeled $t_\PBH{} < t_\mathrm{ev}$ is inconsistent, because PBHs would dominate the universe before evaporating. 
This corresponds to $\beta > \beta_c$ as in \erefs{eq:tPBH}{eq:beta_vs_betac}.  

In the right panel, we vary the parameters associated with the tower, namely the decay rate $\Gamma_1$ and the lightest mass $m_1$, assuming a linear mass spectrum with $m_N = 100 \; m_1$. 
The region labeled $t_\BBN{} < t_R$ is inconsistent, because small values of $\Gamma_1$ correspond to long-lived particles. 
If the particles in the tower decay too late, after BBN they would disrupt the successful predictions of light element abundances, which is in conflict with observational data. 
On the other hand, the region labeled $t_S < t_M$ is inconsistent, because large values of $\Gamma_1$ imply that particles are too unstable such they decay quickly, before they can dominate the energy density of the universe. 
This early decay prevents the stasis period from taking place, which relies on the particles behaving as non-relativistic matter for some time.
This upper boundary on $\Gamma_1$ weakens for a large $m_1$ because the entire tower is also heavier, and closer to the initial PBH temperature. 
This increases the relic abundance of the tower, allowing it to dominate the universe's energy density more quickly. 
As a result, even particles with faster decay rates can still support a stasis phase before decaying. 

\section{Cosmological gravitational particle production}
\label{sec:gpp}

Up to this point we have been considering the evaporation of PBHs as a possible channel for populating stasis' tower of particles.  
We now turn our attention to a second process that allows particles to be created through their gravitational interactions alone.  
The phenomenon of CGPP may occur during the epoch of cosmological inflation or soon after inflation \cite{Ford:2021syk,Kolb:2023ydq}.  
Quantum fields that are not conformally coupled to gravity are excited by the expansion of a homogeneous and isotropic universe \cite{Parker:1968mv,Parker:1969au,Parker:1971pt}.  
After inflation is over and these field excitations are inside the horizon, they may be interpreted as particles.  
The resultant abundances only depend upon the particle's spin and mass, since the interactions are only gravitational, as well as the spacetime expansion history $a(t)$.  
In the remainder of this short section, we briefly remark on how CGPP could produce an adequate number of particles for successful stasis, and we emphasize that the power-law relationship $\Omega_l \propto m_l^\alpha$ is distinct from black hole evaporation. 

First we consider spin-$\ssfrac{1}{2}$ particles. 
We use $\dd n = n_k \tfrac{\dd k}{k}$ to denote the number density of gravitationally-produced particles having comoving wavenumber (equivalently comoving momentum) between $k$ and $k + \dd k$.  
Studies of \CGPP{} with spin-$\ssfrac{1}{2}$ particles of mass $m$ reveal that $n_k$ reaches a maximum at $k \sim k_\ast = a_e H_e^{2/3} m^{1/3}$ where $a_e$ and $H_e$ are the scale factor and Hubble scale at the end of inflation \cite{Chung:2011ck} (see \rref{Kolb:2023ydq} for a review). 
Toward smaller $k$, the spectrum falls as $k^2$, and toward large $k$ it falls off exponentially quickly. 
Consequently, most superheavy spin-$\ssfrac{1}{2}$ particles with $m \sim H_e$ have $k \sim a_e H_e$, corresponding to the comoving Hubble scale at the end of inflation. 

Regarding the total density, these studies also reveal that integrating over $k$ yields a total density $n$ that scales with mass as $n \propto m$.  
For small $m \ll H_e$ the \CGPP{} of spin-$\ssfrac{1}{2}$ particles is suppressed, because a massless spin-$\ssfrac{1}{2}$ particle is conformally coupled to gravity.  
For large $m \gtrsim H_e \approx m_\phi$ the \CGPP{} is also suppressed, because kinematic considerations restrict the production of particles with mass $m$ above the inflaton mass $m_\phi$.  
For intermediate $m \approx H_e$, the particles resulting from \CGPP{} are marginally relativistic, since $k \sim a_e H_e \sim a_e m$, and they quickly become non-relativistic after inflation.   

In the end, the late-time abundance of stable spin-$\ssfrac{1}{2}$ particles that arose from \CGPP{} during inflation is given by (see eqs.~77, 78, and fig.~9 of \rref{Kolb:2023ydq})
\bes{
    \frac{\Omega h^2}{0.12} \approx 
    \biggl( \frac{m/H_e}{10^{-1}} \biggr)^2 
    \biggl( \frac{H_e}{10^{12} \, \mathrm{GeV}} \biggr)^2 
    \biggl( \frac{T_\text{\sc rh}}{10^9 \, \mathrm{GeV}} \biggr) 
    \;.
}
Here $m$ is the mass of the particle, $H_e$ is the Hubble scale at the end of inflation, and $T_\text{\sc rh}$ is the reheating temperature at the start of radiation domination. 
For these fiducial parameters, CGPP predicts $\Omega \sim 0.1$, which is so low that the gravitationally-produced particles would only dominate the universe at radiation-matter equality, which is far too late for a phase of stasis to occur without disrupting cosmological observables. 
Since CMB observations impose an upper bound on the energy scale of inflation \cite{Planck:2018jri}, the most optimistic scenario corresponds to $m/H_e \approx 1$, $H_e \approx 10^{14} \, \mathrm{GeV}$, and $T_\text{\sc rh} \approx 10^{15} \, \mathrm{GeV}$, which leads to $\Omega h^2/0.12 \sim 10^{12}$.  
Since the energy density of a non-relativistic population of particles decreases like $\rho \propto a(t)^{-3} \propto z(t)^3$ while the radiation energy scales like $\rho_\mathrm{rad} \propto a(t)^{-4} \propto z(t)^4$, then these particles that are $10^{12}$ more abundant than the dark matter would dominate the cosmological energy budget at a redshift of $\approx 10^{12} z_\mathrm{eq}$ corresponding to a plasma temperature of $10^{12} T_\mathrm{eq} \approx 1 \, \mathrm{TeV}$. 
If these particles decay soon after they dominate, then there could be a phase of stasis domination during the electroweak epoch. 
We discuss this further below after briefly turning to spin-$0$ and spin-$1$ particles.  

For both spin-$0$ and spin-$1$ particles, \CGPP{} gives rise to a cosmological energy fraction that is parametrically given by (see eq.~78 of \rref{Kolb:2023ydq})
\bes{
    \frac{\Omega h^2}{0.12} \approx 
    \mathrm{min}\begin{cases}
    \bigl( \frac{H_e}{10^{14} \, \mathrm{GeV}} \bigr)^2 
    \bigl( \frac{T_\text{\sc rh}}{10^2 \, \mathrm{GeV}} \bigr) 
    \\ 
    \bigl( \frac{H_e}{10^{14} \, \mathrm{GeV}} \bigr)^2 
    \bigl( \frac{m}{10^{-5} \, \mathrm{eV}} \bigr)^{1/2} 
    \end{cases}
    \;.
}
Once again, for these fiducial parameters we have $\Omega \sim 0.1$, which corresponds to a tiny population of particles that only dominates at radiation-matter equality.  
Taking instead $T_\text{\sc rh} = 10^{15} \, \mathrm{GeV}$ lifts the first case by a factor of $10^{13}$, and taking $m = H_e = 10^{14} \, \mathrm{GeV}$ lifts the second case by a factor of $10^{14}$.  
So under the most optimistic scenario, $\Omega h^2 / 0.12 \sim 10^{13}$, which is a factor of $\sim 10$ larger than \CGPP{} for spin-$\ssfrac{1}{2}$ particles. 

The conclusion is that, with favorable assumptions about the particle mass ($m_l \approx H_e$), the energy scale of inflation ($H_e \approx 10^{14} \, \mathrm{GeV}$), and the efficiency of reheating ($T_\text{\sc rh} \approx 10^{15} \, \mathrm{GeV}$), the phenomenon of \CGPP{} during inflation would create a sufficiently large population of non-relativistic particles to dominate over the radiation in the early universe.  
As these particles decay to radiation, they could drive a phase of matter-radiation stasis.  
However, since the initial abundance cannot be made arbitrarily large, the stasis phase would have to occur at a plasma temperature below $T \approx 1 \, \mathrm{TeV}$, which is roughly the scale of the electroweak phase transition $T \approx 100 \, \mathrm{GeV}$. 
Conversely, if $H_e$ or $T_\text{\sc rh}$ are too small, then $\Omega h^2$ is also reduced, and it would become impossible to achieve stasis before nucleosynthesis. 
It is interesting to note that \CGPP{} predicts
\begin{equation}\label{eq:alpha_CGPP}
    \Omega_l \propto m_l^\alpha \quad \text{with} \quad \alpha = 2 \ \ \text{(for spin-$\ssfrac{1}{2}$)} \quad \text{or} \quad \alpha = 0, \ 1/2 \ \ \text{(for spin-$0$ and -$1$)}
    \;,
\end{equation}
which differs from $\alpha = \pm 1$ for PBH evaporation \eqref{eq:alpha_PBH}.
For successful stasis, the exponent $\alpha$ must obey $-1/\delta < \alpha \leq \gamma/2 - 1/\delta$ as in \eref{eq:constraint}.  
For our fiducial choices of $\gamma = 5$ and $\delta = 1$, these inequalities are $-1 < \alpha \leq 3/2$, which is compatible with \CGPP{} of spin-0 and spin-1 particles, but incompatible with \CGPP{} of spin-$\ssfrac{1}{2}$ particles.  
However, even $\alpha = 2$ is allowed for $\gamma = 7$.

\section{Summary and conclusion}
\label{sec:conclusion}

In this work we have investigated whether gravitational interactions alone are sufficient to provide the initial particle abundances that are needed for a phase of cosmological stasis to occur in the early universe.  
Following earlier work on stasis, we consider a tower of particles (with species label $l$) having masses $m_l$ and rates $\Gamma_l$ to decay outside the tower into radiation.  
We calculate the abundances $\Omega_l$ of gravitationally-produced particles on each level of the tower in two scenarios: PBH evaporation and CGPP during inflation.  
Gravitational interactions offer a generic and (to some extent) unavoidable channel for seeding stasis. 

The main focus of our work entails a study of particle production via PBH evaporation.  
We calculate the spectrum of radiated particles by solving a Boltzmann equation, where the population of PBHs enters as a source term.  
We discuss how the spectrum develops over the black hole's lifetime, and we explain features that appear in the final spectrum.  
For particles that are light compared to the initial black hole temperature ($m_l < T_0$) we find that they are produced with density $n_l \propto m_l^0$; whereas, particles that are heavy ($m_l > T_0$) have $n_l \propto m_l^{-2}$.  
Although the heavy particles are already non-relativistic soon after the black holes fully evaporate, the light particles require a period of cosmological redshifting to become non-relativistic.  
However, eventually the tower of non-relativistic particles comes to dominate the cosmological energy budget, and we find that $\Omega_l \propto m_l^1$ for light particles ($m_l < T_0$) and $\Omega_l \propto m_l^{-1}$ for heavy particles ($m_l > T_0$); see \eref{eq:alpha_PBH}.  
These scalings correspond to $\alpha = \pm 1$ in the stasis parametrization $\Omega_l = (m_l / m_1)^\alpha \Omega_1$ from \eref{eq:Omegal_stasis}. 
The exponent $\alpha$ is constrained by \eref{eq:constraint} to be $-1 < \alpha \leq 3/2$ for $\delta = 1$ and $\gamma = 5$.  
We conclude that PBH evaporation into light particles ($m_l < T_0$) yields the desired scaling ($\alpha = 1$) for stasis, and evaporation into heavy particles ($m_l > T_0$) yields a scaling ($\alpha = -1$) that is also suitable for stasis provided that $0<\delta<1$.  
Finally, we verify that the decay rates $\Gamma_l$ can be chosen such that an extended phase of stasis takes place and the system returns to radiation domination at a temperature $T > 5 \, \mathrm{MeV}$ to avoid disrupting the predictions of big bang nucleosynthesis and neutrino freeze out.  

We also consider CGPP during inflation as a way of preparing the requisite initial conditions for stasis.  
By drawing upon calculations in the literature, we discuss \CGPP{} for spin-$\ssfrac{1}{2}$, spin-0, and spin-1 particles.  
We find that for favorable assumptions about the particle mass ($m_l \approx H_e$), the energy scale of inflation ($H_e \approx 10^{14} \, \mathrm{GeV}$), and the efficiency of reheating ($T_\text{\sc rh} \approx 10^{15} \, \mathrm{GeV}$), \CGPP{} could generate a population of non-relativistic particles that would dominate over the radiation in the early universe as early as $T \approx 1 \, \mathrm{TeV}$.  
Since the cosmological expansion history is almost unconstrained before nucleosynthesis and neutrino decoupling at $T \approx 5 \, \mathrm{MeV}$, this would leave ample time for stasis.  
Moreover this estimate further motivates a study of electroweak phase transition dynamics in a stasis-dominated universe.  
Stasis may leave an observable impacts on the phase transition's cosmological relics including the stochastic gravitational wave background, the primordial magnetic field, and the baryon asymmetry of the universe (\ie{}, baryogenesis).  
We find that the abundance $\Omega_l$ scales as a power-law of the mass $\Omega_l \propto m_l^\alpha$ with $\alpha = 2$ for \CGPP{} of spin-$\ssfrac{1}{2}$ particles and $\alpha = 0$ or $1/2$ for \CGPP{} of bosons; see \eref{eq:alpha_CGPP}.  
It is interesting that these exponents are distinct from particle production via PBH evaporation \eqref{eq:alpha_PBH}, and they can still be compatible with the stasis constraint \eqref{eq:constraint} depending on $\delta$ and $\gamma$.  

Cosmological stasis is a phenomenologically rich example of a modified cosmic expansion history, with potentially unique observable signatures.  
The matter-radiation stasis, which has been the focus of our work, requires a tower of massive and unstable particles.  
This assumption is arguably generic, particularly in the context of theories with compactified extra dimensions that admit spectra of resonances, similar to the Kaluza-Klein tower.  
However, stasis also requires an initial condition such that the particle tower is populated with abundances having a power-law relation to the mass $\Omega_l \propto m_l^\alpha$ and satisfying $-1/\delta < \alpha < \gamma/2 - 1/\delta$.  
Our work has sought to address the question of whether or not this assumption is also generic.  
Relying only on gravitational interactions, we show that the tower can be filled in the necessary fashion through either the evaporation of PBHs or the phenomenon of CGPP during inflation.  
We conclude that the initial conditions are reasonably generic after all.  

 
\acknowledgments
We are grateful to Brooks Thomas for providing valuable feedback on a draft of this article.  The authors are grateful to the Pittsburgh Particle Physics, Astrophysics, and Cosmology Center (PITT PACC) at the University of Pittsburgh for hosting a stimulating workshop on non-standard cosmological epochs where this work began. This material is based upon work supported (in part: A.J.L. and M.V.) by the National Science Foundation under Grant No.~PHY-2412797.  The work of B.S.E. is supported in part by DOE Grant DE-SC-0022021.

\bibliographystyle{JHEP}
\bibliography{refs}
\end{document}